\documentclass[letterpaper,twocolumn,10pt]{article}
\usepackage{usenix}

\usepackage{booktabs}
\usepackage{subcaption}
\usepackage{algorithm}
\usepackage{algorithmic}
\usepackage{listings}
\usepackage{balance}
\usepackage{multirow}
\usepackage{graphicx}
\usepackage[most]{tcolorbox}
\usepackage{tabularx}
\usepackage{placeins}
\tcbuselibrary{listings,skins,breakable}

\definecolor{ruleaction}{RGB}{200,0,0}
\definecolor{ruleheader}{RGB}{0,130,0}
\definecolor{ruleoption}{RGB}{0,0,180}

\newcommand{\system}{\textsc{RuleAutoPilot}}

\definecolor{promptHeader}{RGB}{118,124,130}
\definecolor{promptBody}{RGB}{240,246,255}
\definecolor{promptBorder}{RGB}{70,105,135}

\lstdefinestyle{promptbox}{
  basicstyle=\ttfamily\scriptsize,
  breaklines=true,
  breakatwhitespace=false,
  columns=fullflexible,
  keepspaces=true,
  showstringspaces=false,
  frame=none,
  backgroundcolor=\color{promptBody},
  aboveskip=0pt,
  belowskip=0pt
}

\lstdefinestyle{suricatarule}{
  basicstyle=\ttfamily\scriptsize,
  breaklines=true,
  frame=single,
  backgroundcolor=\color{gray!10},
  numbers=none,
  captionpos=b,
}

\newtcblisting{prompttemplate}[2][]{
  enhanced,
  breakable,
  width=\columnwidth,
  colback=promptBody,
  colframe=promptBorder,
  coltitle=white,
  colbacktitle=promptHeader,
  title=\textbf{#2},
  fonttitle=\ttfamily\scriptsize,
  boxrule=0.5pt,
  arc=1pt,
  left=4pt,
  right=4pt,
  top=4pt,
  bottom=4pt,
  before skip=4pt,
  after skip=6pt,
  listing only,
  listing engine=listings,
  listing options={
    basicstyle=\ttfamily\tiny,
    breaklines=true,
    breakatwhitespace=false,
    columns=fullflexible,
    keepspaces=true,
    showstringspaces=false,
    frame=none,
    backgroundcolor=\color{promptBody},
    aboveskip=0pt,
    belowskip=0pt
  },
  #1
}

\begin{document}

\date{}

\title{\Large \bf \system: Synthesizing Deployable Suricata Rules from Network Traffic}

\author{
{\rm Mughees Ur Rehman}\\
Purdue University\\
mughees@purdue.edu
\and
{\rm Aritran Piplai}\\
University of Texas at El Paso\\
apiplai@utep.edu
\and
{\rm Murat Kantarcioglu}\\
Virginia Tech\\
muratk@vt.edu
}

\maketitle

\begin{abstract}
Rule-based Intrusion Detection Systems (IDS) such as Suricata are central to
network security, yet crafting effective detection rules demands deep expert
knowledge and cannot keep pace with emerging threats. Existing LLM-based approaches can reduce analyst effort, but they either rely on curated threat intelligence that is produced only after the underlying traffic artifacts already exist, or they require costly LLM use without sufficient quality control.
We present \system{}, an end-to-end agentic framework that generates
deployable Suricata rules directly from malware network traffic, with no
prior threat intelligence required. A key challenge is noise: network traffic captures often contain a small amount of security-relevant traffic mixed with large volumes of background traffic, which reduces LLM reasoning quality and increases cost. \system{} addresses this challenge with a Benign Traffic Fingerprinting stage that removes known benign background flows before LLM processing. Rules that fail
syntax checks, do not trigger on the source traffic, or generate false
positives on a benign corpus are automatically repaired using structured
feedback. Across 1,296 malware PCAPs, execution-grounded verification raises rule
quality (F1) from 0.443 to 0.539. On a stratified 200-PCAP subset,
\system{} on the open-weight \texttt{gpt-oss-120b} reaches near-frontier
quality, 0.524 F1 against Claude Opus~5 under Claude Code's 0.623, at
52$\times$ lower billed-token cost. Swapping only the backbone to Claude
Opus~5, \system{} surpasses Claude Code outright, 0.656 F1 against 0.623,
at 40$\times$ fewer tokens. A stronger backbone raises \system{}'s own
ceiling, but at the same backbone, our scaffold still outperforms Claude
Code's, showing the scaffold contributes independently of the backbone.
\end{abstract}
\section{Introduction}
\label{sec:introduction}

Signature-based Intrusion Detection Systems (IDS) such as
Suricata~\cite{suricata} and Snort~\cite{snort} remain widely deployed
components of operational network defense~\cite{albin2012realistic,sommestad2021variables}.
They inspect network traffic against databases of \emph{rules} (structured signatures that encode protocol-level characteristics of known threats) and raise alerts or block connections when matches are found. Suricata is well suited for high-throughput deployments, partly due to its multi-threaded design~\cite{tayyebi2018comparative,waleed2022which}.

Community-maintained rulesets such as Emerging Threats Open
(ET~Open)~\cite{etopen} provide curated signatures for known malicious behavior. Suricata documentation describes ET~Open as a free ruleset and useful
signature reference~\cite{suricata_rules_intro}, while prior work shows that
NIDS effectiveness depends heavily on rule quality~\cite{sommestad2021variables}.

Despite their effectiveness, signature-based IDS are constrained by a persistent rule creation bottleneck. Writing a high-quality Suricata rule requires substantial analyst effort: analysts must inspect traffic, infer the underlying behavior, and design specific rules that provide useful coverage without creating noisy alerts~\cite{teuwen2025ruling,diazverdejo2022detection}. The resulting rule must then be tested to ensure that it loads correctly, detects the intended malware traffic, and avoids excessive false positives~\cite{moreno2025leveraging}. This process is difficult to scale: AV-TEST reports that more than 450,000 new malware and potentially unwanted application samples are registered each day~\cite{avtest2024}.

Recent work has explored Large Language Models (LLMs) for automating IDS rule generation. For instance, RuleMaster+~\cite{rulemaster2025} fine-tunes an LLM on instruction datasets derived from proof-of-concept exploit descriptions. FALCON~\cite{falcon2025} generates Snort and YARA rules from Cyber Threat Intelligence (CTI) reports using an agentic framework. Moreno et al.~\cite{moreno2025leveraging} study zero-shot and few-shot prompting for Suricata rule generation from pre-segmented malicious flows in ICS/SCADA settings, with syntax-only validation.

While promising, these approaches still assume that rule generation starts from an already processed security artifact, such as a PoC write-up, CTI report, or analyst-labeled malicious network segment. This assumption moves the hardest part of the workflow outside the system: before a rule can be generated, someone must first analyze the raw evidence and identify the security-relevant behavior. In early-stage incidents, however, security analysts may only have a suspicious binary or a malware execution trace. The malicious behavior has not yet been summarized into CTI, distilled into curated indicators, or separated from benign background traffic.

This gap is not a historical artifact. On 3~December 2025, researchers publicly disclosed CVE-2025-55182 (``React2Shell''), a CVSS~10.0 pre-authentication remote code execution flaw in the Flight protocol used by React Server Components and Next.js~\cite{unit42_react2shell,datadog_react2shell}. No active exploitation had been reported at disclosure; within five days, post-exploitation activity was observed on vulnerable hosts~\cite{unit42_react2shell}. Among the concrete detection artifacts circulated by defenders during that window were a packet capture of exploit traffic together with Suricata and Snort signatures derived from it~\cite{vulncheck_react2shell}, which is precisely the traffic-to-rule workflow this paper automates. More broadly, CTI gathering, analysis, and reporting can take days or weeks~\cite{shah2024cti}.  

Another challenge is TLS encryption, which leaves a rule only the packet headers and the handshake to work with. The problem is new and consequential, since traditional Suricata rules match on payload as well. Yet established rulesets such as ET~Open continue to derive substantial value from plaintext inspection: of the 13,645 rules ET~Open published over the past two years, 36\% target encrypted traffic while 64\% read plaintext. A rule-generation system must therefore produce high-quality rules for both encrypted and unencrypted traffic.

This gap motivates a central research question: \textit{How can we
automatically synthesize deployable Suricata rules directly from network traffic, without relying on curated threat intelligence, manually labeled malicious flows, or a security analyst in the loop?}

To address this problem we propose \system{}, an agentic framework for
end-to-end Suricata rule synthesis from network traces, designed around three operational requirements: generated rules must be syntactically valid, must
trigger on the traffic they were derived from, and must not fire on benign
traffic, because false alarms make a rule undeployable regardless of what it
detects. The first challenge is \emph{noise}. Malware traces contain a few
security-relevant flows among many background ones, and passing all of them to an
LLM raises cost, can exceed the context window, and distracts the model from the
flows that matter. \system{} addresses this with \emph{Benign Traffic
Fingerprinting}, a protocol-aware filtering stage that removes recurring
background flows before synthesis.

The second challenge is \emph{deployability}, and \system{} answers it with a
verification-first design. A synthesis agent identifies security-relevant flows
and writes candidate rules; a verification agent then loads each rule in
Suricata, replays it against the source malware PCAP, and replays it against a
benign corpus. Rules failing any check go to a repair agent that revises them
from structured verifier feedback before re-verification. The pipeline therefore
does not produce plausible rule text but executable rules tested against real
malware and benign traffic.

Building the system makes a second question answerable, and it is the one the
paper is ultimately about. Rule generation needs a model, a prompt, and a
\emph{scaffold}: the code around the model that decides what it sees and what
happens to its output. It is not obvious which of the three determines quality. If capability dominates, the answer is to wait for larger models. If the
scaffold does, an open model that can be run in-house on sensitive
traffic already suffices. We answer this two ways. First, we hold \system{}'s scaffold fixed and vary
the backbone, running it on \texttt{gpt-oss-120b} and on Claude Opus~5, to
measure how much the backbone alone contributes. Second, we hold the
backbone fixed and vary the scaffold: at \texttt{gpt-oss-120b}, we compare
\system{} against two general-purpose coding agents, OpenClaw and Hermes
Agent; at Claude Opus~5, we compare \system{} against Claude Code. Together
these tell us whether the backbone or the scaffold drives rule quality. The
prompt is measured separately, by sweeping six variants inside one
scaffold. These questions structure the paper:

\begin{description}
    \item[\textbf{RQ1:}] How well can an agentic LLM pipeline generate deployable Suricata rules from malware traffic, and how closely do they align with ground-truth rules?

    \item[\textbf{RQ2:}] What determines rule quality, the backbone model or the scaffold around it, and does a general-purpose scaffold suffice?

    \item[\textbf{RQ3:}] When traffic is encrypted, what evidence remains
visible, and how well can Suricata rules still be generated from it?
     
\end{description}

We evaluate \system{} on 1,296 malware PCAPs across 192 malware families, spanning samples from 2014 to 2026, along with 1,172 benign PCAPs across the benign fingerprinting and benign evaluation corpora. The cross-system comparison uses a 200-capture stratified subset drawn from the same corpus.
We make the following core contributions:
\begin{itemize}
    \item \textbf{End-to-end rule generation from network traffic.}
    \system{} generates Suricata rules directly from malware PCAPs, with no CTI report, CVE description, or analyst-labeled flow as input.

    \item \textbf{Benign traffic fingerprinting.}
    We introduce a protocol-aware filtering stage built from 981 benign PCAPs. It removes 88.8\% of background flows from malware PCAPs while retaining 98.0\% of security-relevant flows, reducing average LLM token usage by 1.99$\times$.

    \item \textbf{Execution-grounded validation and repair.}
    Every candidate rule is loaded in Suricata, replayed on the source
    malware PCAP, and replayed on benign traffic, with failures sent to a
    repair agent. Together these raise flow-alignment F1 (how closely a
    generated rule's alerts match the flows an expert-authored rule would
    flag) from 0.443 to 0.539 and cut the benign false-positive rate from
    0.128\% to 0.006\%.

\item \textbf{At a fixed backbone, the scaffold determines rule quality.}
At a fixed \texttt{gpt-oss-120b} backbone, \system{} reaches 0.524
flow-alignment F1 where the same model driving the
OpenClaw~\cite{openclaw} and Hermes~\cite{hermes_agent} coding agents, on
an identical prompt, reaches 0.416 and 0.400. Every system receives the
same benign false-positive control, so the comparison does not reward
\system{} for a stage the baselines lack.

\item \textbf{Frontier-model rule quality at open-model cost.}
Claude Opus~5 under Claude Code~\cite{claude_code} reaches 0.623 F1
against \system{}'s 0.524 on \texttt{gpt-oss-120b}, at 52$\times$ more
billed tokens per capture. A scaffold built for rule generation therefore
comes close to frontier-model quality at a fraction of the cost.

\item \textbf{At equal backbone, our scaffold outperforms Claude Code's.}
Pairing \system{}'s own scaffold with the same Claude Opus~5 backbone that
powers Claude Code surpasses Claude Code's F1 outright, 0.656 against
0.623, while using 40$\times$ fewer tokens. Holding the backbone fixed
isolates the scaffold as the source of the gap.

\end{itemize}

\section{Background}
\label{sec:background}

We provide background on Suricata rule structure and the community-maintained rulesets used in operational deployments.

\subsection{Rule Structure}
\label{sec:bg-suricata}

Suricata analyzes traffic against structured rules to raise alerts~\cite{suricata}. Each rule has an action and a header defining protocol, address, port
scope, and direction. It also has a set of options that define matching
conditions over flows, payloads, and protocol fields, plus metadata for
identification and maintenance.

\begin{figure}[t]
\centering
\setlength{\fboxsep}{4pt}
\fbox{%
\begin{minipage}{0.96\columnwidth}
\scriptsize\ttfamily
\textcolor{red}{alert}
\textcolor{green!50!black}{dns \$HOME\_NET any -> any any}
\textcolor{blue}{( msg:"ET MALWARE Diezen/Sakabota CnC Domain}\\
\textcolor{blue}{Observed in DNS Query";}\\
\textcolor{blue}{\quad dns.query; content:"antivirus-update.top"; nocase; endswith;}\\
\textcolor{blue}{\quad classtype:domain-c2; sid:2029326; rev:2;}\\
\textcolor{blue}{\quad metadata:created\_at 2020\_01\_29, deployment Perimeter,}\\
\textcolor{blue}{\quad confidence High, signature\_severity Major, updated\_at 2020\_01\_29; )}
\end{minipage}%
}
\caption{Rule matching a malware C2 domain in DNS traffic.}
\label{fig:suricata-rule-example}
\end{figure}

Figure~\ref{fig:suricata-rule-example} shows these parts on a C2 domain
rule. \texttt{alert} is the action, and the header restricts matching to
DNS traffic leaving the monitored network. Among the options,
\texttt{dns.query} selects the query field and \texttt{content} looks for
the domain \texttt{antivirus-update.top} within it. The
\texttt{classtype}, \texttt{sid}, \texttt{rev}, and \texttt{metadata}
fields identify and classify the rule.

\subsection{Deep Packet Inspection and Stateful Matching}
\label{sec:bg-matching}

Suricata rules match in two distinct ways, and this paper's results turn on the
difference.

\noindent\textbf{Content matching.} Most rules inspect bytes, either raw
payload or one of the many parsed protocol buffers Suricata exposes,
including \texttt{http.uri}, \texttt{dns.query}, and \texttt{tls.sni}.
This is deep packet inspection, and it requires those bytes to be
readable. Transport encryption hides everything above the TLS record
layer, including HTTP headers, paths, and bodies, while the TLS handshake
itself, including the server name, stays visible before encryption
begins. What a capture exposes therefore bounds what a content rule can
key on (\S\ref{sec:encryption}).

\noindent\textbf{Stateful matching.} A minority of rules express
conditions over accumulated network state rather than a single packet.
\texttt{threshold} and \texttt{detection\_filter} keywords count events
within a time window: ET~Open rule \texttt{sid:2001581}, for instance,
alerts only when one internal host opens at least 70 TCP connections to
port~135 within 60 seconds. \texttt{flowbits} sets and tests named bits
so that one rule can condition on another having matched earlier in the
same flow.

\subsection{Rule Maintenance and Community Rulesets}
\label{sec:bg-maintenance}

Rule authors must balance specificity against generality: broad rules produce false positives, narrow ones miss variants of the same behavior. Community-maintained feeds such as URLhaus and Feodo Tracker publish Suricata-compatible rules for known malicious infrastructure~\cite{urlhaus_ids,feodo_tracker}. These feeds operationalize known indicators quickly, but they are indicator-driven: they match specific URLs, domains, IP addresses, or IP:port pairs, so their coverage degrades when operators rotate infrastructure. Emerging Threats Open (ET~Open)~\cite{etopen} instead provides rules that inspect protocol fields, payload content, and malware-specific network behavior. The Suricata documentation describes it as a free ruleset with a wide range of signature examples, and the ET community publishes frequent updates~\cite{suricata_rules_intro,et_updates}. ET~Open has been adopted as a baseline or comparison ruleset in prior IDS evaluations~\cite{white2013quantitative,cordero2021generating,gridai2025}, which makes it a reliable reference for studying how high-quality signatures capture malicious network behavior. In this paper we use it as our ground truth.

\section{Problem Statement and Threat Model}
\label{sec:problem}

\noindent\textbf{Problem Statement.}
Rule authoring requires expertise in traffic analysis, malware behavior, and rule construction, so manual generation struggles to keep pace. Deployment needs more than plausible rule text: candidates must be executable by Suricata and validated against both the source malware traffic and benign traffic. Given a trace, \system{} produces rules that are syntactically valid, trigger on the source traffic, and stay quiet on benign traffic.

\noindent\textbf{Assumptions.}
We assume that \system{} is given either an existing network traffic capture from malware execution or a malware binary that can be executed in a controlled environment to obtain such traffic. \system{} \textit{does not} require curated CTI reports, CVE write-ups, attacker-IP lists, proof-of-concept descriptions, or malware-family labels as input. In addition, \system{} requires access to a Suricata environment for rule validation and a benign traffic corpus for false-positive testing. We assume that rule generation is performed in a trusted environment.

\noindent\textbf{Adversary Model.}
We consider malware that exhibits network-visible behavior, such as command-and-control communication, staged payload delivery, credential theft, or exfiltration. We assume that this behavior is at least partially expressible in Suricata's rule language. We do not consider an adaptive adversary that observes \system{} and then crafts traffic specifically to evade the generated rules. Analyzing such adversarial adaptation is beyond the scope of this work.

\section{\system{} Architecture}
\label{sec:system}

\begin{figure*}[t]
    \centering
    \includegraphics[width=0.82\textwidth]{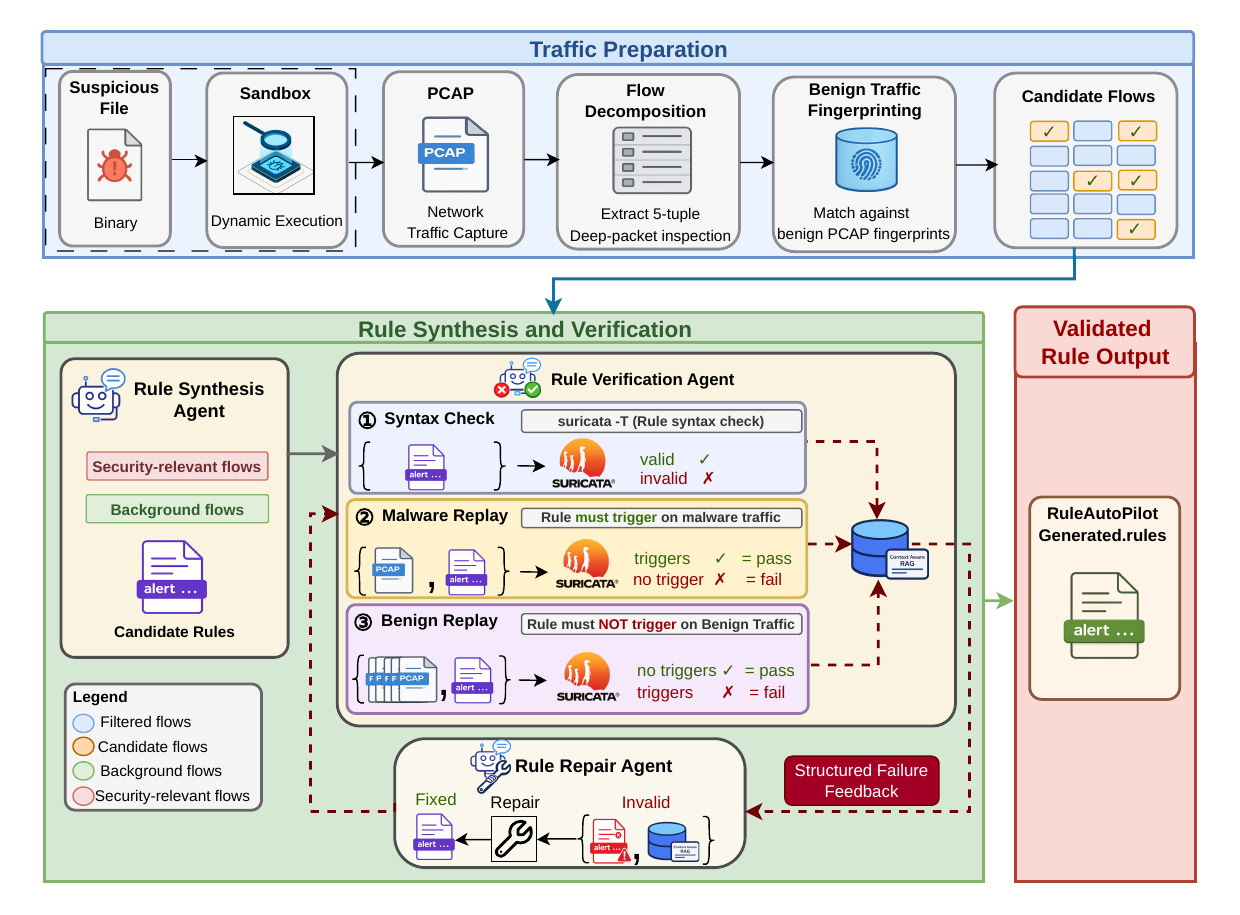}
    \caption{\system{} architecture for traffic preparation, rule synthesis, verification, and repair.}
    \label{fig:architecture}
\end{figure*}

\system{} synthesizes and validates Suricata rules from malware network traces.
It accepts either a suspicious file,
which it executes in a sandbox to capture a PCAP, or an already collected trace.
The PCAP is decomposed into flows enriched with Suricata-compatible protocol
fields, flows matching recurring benign execution patterns are removed, and the
remaining candidate flows are passed to the Rule Synthesis Agent. Rather than
forcing a rule per retained flow, \system{} leaves it to the model to synthesize
rules only where a flow or flow group exposes stable, rule-expressible evidence.
Every generated rule is then validated by Suricata execution, and rules that fail
are repaired from structured feedback.

Four components implement this: \textbf{Benign Traffic Fingerprinting}
(\S\ref{sec:fingerprinting}) removes recurring benign execution patterns; the
\textbf{Rule Synthesis Agent} (\S\ref{sec:synthesis}) separates security-relevant
from background flows and writes rules for the former; the \textbf{Rule
Verification Agent} (\S\ref{sec:verification}) checks syntax, malware triggering,
and benign false positives by execution; and the \textbf{Rule Repair Agent}
(\S\ref{sec:fixer}) revises failures from structured feedback and a curated
repair knowledge base. Figure~\ref{fig:architecture} shows the four.

\subsection{Benign Traffic Fingerprinting}
\label{sec:fingerprinting}

Malware traces often carry hundreds to thousands of flows, most of them benign background activity: DNS lookups, update checks, time synchronization, certificate validation. To reduce this noise, \system{} begins with \emph{Benign Traffic
Fingerprinting}. We take 981 benign samples from a DIKE-derived
corpus~\cite{dike}, confirm their status with VirusTotal~\cite{virustotal}, and
execute them in Cuckoo Sandbox~\cite{cuckoo} to collect PCAPs. From these we
extract recurring protocol-level patterns that characterize normal execution:
DNS query names, HTTP hosts, URIs and User-Agents, TLS SNI values and stable
handshake identifiers, and recurring artifacts such as NTP destinations, OCSP
behavior, and broadcast traffic. Table~\ref{tab:benign-fingerprint-categories}
lists the categories. When processing a PCAP, \system{} extracts the same features per flow and removes
library matches, retaining the rest as candidates. The filter is deliberately
conservative because a discarded security-relevant flow cannot contribute to any
rule.

The filter removes 88.8\% of background flows while discarding 2.0\% of security-relevant ones; \S\ref{sec:results-fingerprinting} and Table~\ref{tab:benign-filter-impact} quantify the effect.

\subsection{Rule Synthesis Agent}
\label{sec:synthesis}

The Rule Synthesis Agent receives the candidate flows surviving benign
fingerprinting and decides which contain behavior suitable for rule
generation. This is narrower than malicious/benign classification: a flow
is useful only if it exposes stable, discriminative features expressible
in Suricata's rule language. Working from the structured flow
representation rather than raw packets, the agent looks for two kinds of
evidence. The first is a protocol field that names a specific indicator,
such as a DNS query, HTTP path, host header, or TLS SNI value. The second
is behavior visible only across multiple flows, such as a timestamped
connection timeline, per-destination connection counts and inter-arrival
times, or the number of distinct destinations a host contacts. A flow
lacking either kind of evidence, or carrying only routine activity,
unstable artifacts, or indicators too broad to be reliable, is treated as
background; ambiguous evidence resolves toward retention.

A triage step examines each capture and decides which call to invoke:
captures with specific-indicator evidence go to the content call, captures
with cross-flow behavioral evidence go to the stateful specialist call,
and captures with both kinds of evidence go to each. 

\noindent\textbf{Deep packet inspection.} The content call writes rules
that match payload bytes or a parsed protocol buffer with an appropriate
protocol and direction. It avoids rules resting only on incidental IP
addresses or ports, and prefers protocol-aware conditions specific enough
to avoid benign matches. Where several flows share a pattern, it writes
one rule capturing their invariant features rather than one per flow; the
rule need not reproduce the exact ET~Open rule, since behaviorally
equivalent rules may differ syntactically.

\noindent\textbf{Stateful rule generation.} The specialist call carries
roughly 600 tokens of stateful-only instruction and none of the
content-match demonstrations, so the examples that make content rules
precise do not pull it back toward writing one. Rules it produces
typically carry a content anchor or a specific destination alongside
their rate condition, following the same logic as ET~Open's own
rate-based signatures, which almost always pair a threshold with a
content match rather than firing on volume alone.

Every produced rule is combined into a single set of
candidate rules, which then goes to the Rule Verification Agent. We evaluate six prompting variants for the content call
(\S\ref{sec:setup}); templates are in
Figures~\ref{fig:prompt-zeroshot}--\ref{fig:prompt-fewshot}.

\subsection{Rule Verification Agent}
\label{sec:verification}

The Rule Verification Agent determines whether a candidate rule is
deployable by running it with the Suricata engine. A rule is accepted
only if it passes three checks in sequence, and the failure label and
execution context of the first failed check become the repair agent's
input.

\noindent\textbf{Syntax.} Each candidate is loaded with
\texttt{suricata -T}, which catches unsupported keywords, invalid
headers or directions, a missing \texttt{sid}, misused sticky buffers,
and invalid PCRE. The parser error is recorded verbatim.

\noindent\textbf{Malware trigger.} Survivors are replayed against the
source malware PCAP. A trigger does not prove the matched flow
malicious; it confirms the rule captures behavior present in the
execution. Failures, most often direction mismatches or overly
restrictive content matches, are recorded with the flow context the
rule was expected to match. Rules that use the \texttt{flowbits}
keyword are an exception, since they belong to the stateful rule class
and come in pairs: one sets a named bit when it matches and raises no
alert itself, and a second fires only if that bit is already set.
Verifying the setter alone would record a trigger failure for a rule
that is working correctly, so this check is applied to the pair as a
unit, and a rule that tests a bit is rejected unless the rule setting
it is also present.

\noindent\textbf{Benign false positive.} Rules that trigger are
replayed against a benign PCAP corpus, where firing means the rule is
too broad. The verifier records both the background flows that fired
and the malware flows the rule matched, which is what lets the repair
agent narrow it while preserving the intended match.

A rule passing all three checks is accepted.

\subsection{Rule Repair Agent}
\label{sec:fixer}

When verification rejects a candidate, the \emph{Rule Repair Agent}
attempts repair within a fixed budget ($K = 3$ throughout). Which
strategy applies depends on the failure and on which synthesis call
produced the rule.

\noindent\textbf{Deterministic rewrites.} For failure patterns with a
precise programmatic fix, a rewrite is applied before any LLM call, such
as swapping source and destination variables when the direction
contradicts the header. A rewritten rule that passes verification is
accepted with no LLM call; Table~\ref{tab:deterministic-rewrites} lists
all 17.

\noindent\textbf{Context-aware retrieval-augmented repair.} Otherwise
the agent uses the verifier's structured feedback, comprising failure
type, verifier output, and flow context, to retrieve a matching entry
from a curated knowledge base of 42 error classes and build a
failure-specific prompt. Syntax prompts carry the parser error and the
retrieved guidance. Trigger prompts carry the flows the rule was
expected to match. False-positive prompts carry both the background
flows that fired and the malware flows that matched. The knowledge base
was curated from documented Suricata constraints, parser-error patterns,
and practitioner repair
workflows~\cite{suricataRuleFormat,suricataPayloadKeywords,suricataDnsKeywords,emergingThreatsFpWorkflow}.
Retrieval keeps each prompt narrow rather than exposing the LLM to
unrelated advice. Tables~\ref{tab:error-patterns-syntax}--\ref{tab:llm-rewrites}
and Figures~\ref{fig:prompt-fixer-system}--\ref{fig:prompt-false-positive-feedback}
give the full taxonomy, strategy coverage, and templates.

\noindent\textbf{Stateful bundles.} When a stateful bundle fails
verification, the rule itself is already valid Suricata syntax; the
problem is what it is detecting, not how it is written. So the
deterministic rewrites used for other rules do not apply here. One
rewrite in particular is deliberately withheld: stripping the
\texttt{threshold} clause, since that clause is the very condition the
rule exists to enforce, and removing it would defeat the rule's purpose. Instead, the repair agent draws on seven behavioral failure patterns
specific to bundles. These include a rate-based rule with no content
match to anchor it and a stateful bit that a rule checks for but that no rule
ever sets, among others.

After each attempt the verifier re-runs the full sequence. Rules still
failing after $K$ iterations are excluded from the accepted rule set.

\section{Dataset and Ground Truth Construction}
\label{sec:dataset}

A core challenge in evaluating automated IDS rule generation is the lack of standardized datasets that pair network traffic with rules. We therefore construct malware and benign traffic corpora from real execution artifacts, prioritizing real-world samples over simulated traces.

\noindent\textbf{Malware corpus.}
We collect Windows PE malware from MalwareBazaar~\cite{malwarebazaar}, ANY.RUN~\cite{anyrun}, and Triage~\cite{triage}, executing binaries in Cuckoo Sandbox~\cite{cuckoo} and using Triage PCAPs directly. Each sample's VirusTotal report~\cite{virustotal} supplies \texttt{first\_submission\_date}, giving a 2014--2026 span (Table~\ref{tab:malware-temporal-distribution}), and family labels are normalized with ClarAVy~\cite{claravy}. The benchmark holds 1,296 PCAPs across 192 families, with 15 unlabeled PCAPs treated as singletons (Table~\ref{tab:malware-family-size-distribution}).

\noindent\textbf{Benign corpora.}
We use two benign corpora, kept disjoint so that fingerprinting data never overlaps abstention-evaluation data. The first is 981 DIKE-derived PCAPs~\cite{dike}, used to build the fingerprint library and, in subsets, for benign replay and false-positive testing. The second is a 191-PCAP external benchmark from the AsiaCCS 2021 benign binaries dataset~\cite{Lucas21Malware}, all VirusTotal-verified benign, used to test whether the pipeline wrongly generates rules where it should generate none.

\noindent\textbf{Rules-PCAP ground truth.}
Running Suricata with ET~Open on each PCAP yields the ground-truth rules it triggers. Every malware PCAP triggers at least one; no benign PCAP triggers any. The triggered rules span \texttt{created\_at} dates from 2010 to 2026 (Table~\ref{tab:rule-temporal-distribution}), so both older and recent signatures remain relevant here.

\noindent\textbf{Flow decomposition.}
\label{sec:flow-decomposition}
We decompose each PCAP into 5-tuple flows and enrich each with Suricata-compatible protocol fields extracted by \texttt{tshark}, selected per protocol using heuristics from the ET~Open ruleset. Fields without Suricata equivalents are dropped and \texttt{tshark} names remapped where necessary; Table~\ref{tab:protocol-fields} lists the full set. A flow is labeled \emph{security-relevant} if it triggers at least one ET~Open rule and \emph{background} otherwise, which yields a dataset-level mapping from each security-relevant flow to its triggered rule(s).

\noindent\textbf{Stratified subset for cross-system comparison.}
\label{sec:strata}
Running four additional systems over all 1,296 captures is not affordable for the
frontier model, so we draw a 200-capture stratified subset spanning 104
families (Table~\ref{tab:strata200}). Strata follow what the ground truth
requires: a content match on an indicator visible in cleartext, on a partly
encrypted trace, on a fully encrypted trace, or a rate or ordered sequence no
single-packet rule expresses, which we call \emph{stateful}. That last stratum
is 81 of 1,296 captures (6.2\%) and is oversampled to 40 of 200 for statistical
power, so per-stratum scores are reported separately rather than pooled. The
encrypted stratum draws its 40 captures from only 15 families and is more
family-concentrated than the others.

\begin{table}[t]
\centering\small
\caption{Composition of the 200-capture cross-system evaluation subset.}
\label{tab:strata200}
\begin{tabular}{@{}lrr@{}}
\toprule
Ground-truth type & PCAPs & Families \\
\midrule
Content, cleartext & 60 & 60 \\
Content, mixed     & 60 & 52 \\
Content, encrypted & 40 & 15 \\
Stateful         & 40 & 18 \\
\midrule
Total              & 200 & 104 \\
\bottomrule
\end{tabular}
\end{table}

\noindent\textbf{Dataset summary.}
Table~\ref{tab:dataset-stats} summarizes the malware and benign datasets used in our evaluation.

\begin{table}[t]
  \centering
  \caption{Dataset summary for the rule-generation benchmark and benign corpora.}
  \label{tab:dataset-stats}
  \small
  \begin{tabular}{@{}lr@{}}
    \toprule
    \textbf{Metric} & \textbf{Value} \\
    \midrule
    \multicolumn{2}{@{}l}{\textit{Malware corpus}} \\[2pt]
    Malware benchmark PCAPs                & 1,296 \\
    Unique malware families                &   192 \\
    Single-PCAP (no-family) PCAPs          &    15 \\
    Security-relevant flows                & 6,937 \\
    Background flows                       & 124,850 \\
    Unique ET~Open rules triggered                &   617 \\
    \midrule
    \multicolumn{2}{@{}l}{\textit{Benign corpora}} \\[2pt]
    DIKE-derived benign PCAPs              &   981 \\
    External benign-execution benchmark    &   191 \\
    \bottomrule
  \end{tabular}
\end{table}


\section{Evaluation}
\label{sec:evalframework}

We evaluate \system{} through the three research questions of
\S\ref{sec:introduction}.

\subsection{Metrics}
\label{sec:metrics}

\noindent\textbf{Flow classification, syntax pass, and trigger rate.}
Flow classification is the binary task of separating security-relevant flows from background, using the labels of \S\ref{sec:flow-decomposition}. Syntax pass rate is the fraction of generated rules that load in Suricata after repair; malware trigger rate is the fraction of accepted rules that fire on the source PCAP, the minimum behavioral requirement for deployment.

\begin{figure}[t]
    \centering
    \includegraphics[width=\columnwidth]{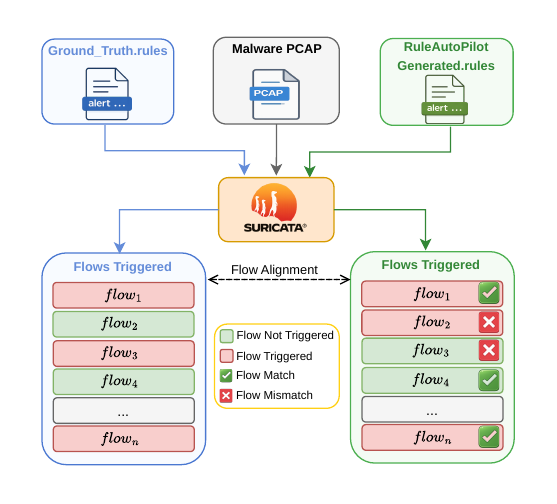}
    \caption{Flow-level comparison of ground-truth and generated rules on the same malware PCAP. FAS quantifies how closely generated-rule alerts align with ground-truth alerts at the flow level.}
    \label{fig:fas}
\end{figure}

\noindent\textbf{Flow Alignment Score (FAS).}
FAS measures how closely a generated rule matches the flow-level alert behavior of the corresponding expert-authored ET~Open rule, as illustrated in Figure~\ref{fig:fas}.
We execute both rules on the same malware PCAP and compare the alerted
flow sets. FAS precision is the fraction of generated-rule alerts that match on ground-truth security-relevant flows; FAS recall is the fraction of ground-truth security-relevant flows covered by the generated rules.

\noindent\textbf{Coverage.}
A system emitting no rules for a capture has produced no detection, so we score
it zero there. Because that alone understates deliberate abstention, we also
report \emph{coverage}, the fraction of captures answered. Where coverage falls
well short of 100\%, we give precision over answered captures alongside the
pooled figure, since the two then measure different things.

\noindent\textbf{Over-fire ratio.}
Over-fire is alerted flows divided by ground-truth security-relevant flows,
pooled across captures; 1.0 means the system raises as many flow-level alerts as
the reference ruleset. Precision cannot separate a rule firing twice too often
from one firing fifty times too often, and analyst load tracks the latter.

\noindent\textbf{False positive rate (FPR).}
FPR is the fraction of benign PCAPs triggered by the generated rule set: the number of benign PCAPs with at least one alert divided by the total benign PCAPs tested. Lower FPR means fewer false alerts on benign traffic.

\subsection{Experimental Setup}
\label{sec:setup}

The primary backbone is \texttt{gpt-oss-120b}, evaluated under six prompting
strategies (zero-shot, CoT, few-shot-1/3/5, CoT+FS-3); we use few-shot-5
throughout unless noted, and additionally evaluate GPT-5.5 on a stratified
100-PCAP subset. Benign replay for verification
and false-positive testing draws on disjoint subsets of the 981-PCAP DIKE-derived
corpus, with an external 191-PCAP benchmark~\cite{Lucas21Malware} reserved for
benign-only abstention. For RQ2, we compare \system{} against four additional
systems on the 200-capture stratified subset of \S\ref{sec:strata} under a
uniform benign false-positive control, with all baseline agents receiving a
byte-identical prompt (verified by hash) to isolate scaffold and model effects.
Full model parameters, prompting details, the comparison-system harness, the
false-positive control procedure, token-accounting methodology, and the
Suricata validation environment are given in Appendix~\ref{app:setup-details}.

\subsection{RQ1: End-to-End Rule Generation}
\label{sec:results-known}

RQ1 asks how well \system{} generates deployable rules from raw traffic. We answer this along three axes: i) rule quality on malware PCAPs, measured by flow-classification F1, syntax pass rate, malware trigger rate, and the Flow Alignment Score (FAS); ii) whether the pipeline correctly
abstains on benign traffic rather than manufacturing rules where none are needed; and iii) the generalizability of generated rules to unseen samples within and across malware families, rather than overfitting to their source PCAP.

\noindent\textbf{Evaluation on malware PCAPs.}
We try different in-context learning configurations, including zero-shot,
chain-of-thought, and few-shot prompting, and use few-shot-5 as our default
throughout. We report macro-averaged scores, weighting each PCAP equally
regardless of its flow count, since a corpus dominated by a few flow-heavy
captures would otherwise obscure how the pipeline performs on a typical
sample. Under this scoring, \system{} reaches FAS-F1 0.54 across the 1,296
malware PCAPs, with an 85.3\% syntax pass rate, an 81.1\% malware trigger
rate, and a 0.006\% benign FPR. Prompt choice barely changes this: across the
three configurations (Table~\ref{tab:main-prompt-comparison}), FAS-F1 moves
by only 0.018.

\noindent\textbf{Performance on malware unseen during training.}
Because \system{} needs no labels or pre-written IoCs, generated rules should
not depend on memorized training content from the model. To check this, we
evaluate on post-cutoff malware: \texttt{gpt-oss-120b}'s training cutoff is
June~2024, and 1,116 of the 1,296 malware PCAPs postdate it. These PCAPs
trigger 480 unique ET~Open rules, 214 of which were created after the
cutoff, so the model cannot simply be recalling them.
On this subset, \system{} with few-shot-5 reaches a FAS-F1 of 0.533, closely
tracking the corpus-wide figure of 0.54, with an 82.1\% trigger rate and an
85.1\% syntax pass rate. The pipeline leans toward recall over precision
here, which is arguably the right trade in a zero-day setting: missing a
security-relevant flow is more costly than surfacing an extra candidate for
an analyst to dismiss. Benign FPR stays low at 0.0072\%, so this recall bias
does not flood analysts with rules on benign traffic. These results show
that \system{} performs well on malware the model could not have
memorized, so its rule quality comes from reasoning over observed traffic
rather than recall.

\begin{table*}[htbp]
 \caption{End-to-end performance for zero-shot, CoT, and few-shot-5, macro-averaged over 1,296 malware PCAPs (full verification, \texttt{gpt-oss-120b}). The full six-variant sweep is in Appendix~\ref{app:macro-prompt-comparison}.}
  \label{tab:main-prompt-comparison}
  \small
  \centering
  \begin{tabular}{@{}lccc|cccccc@{}}
    \toprule
    & \multicolumn{3}{c|}{\textbf{Flow Classification}}
    & \multicolumn{6}{c}{\textbf{Rule Quality}} \\
    \cmidrule(lr){2-4}
    \cmidrule(lr){5-10}
    \textbf{Variant}
    & \textbf{P} & \textbf{R} & \textbf{F1}
    & \textbf{FAS-P} & \textbf{FAS-R} & \textbf{FAS-F1}
    & \textbf{Syntax} & \textbf{Trigger} & \textbf{FPR} \\
    \midrule
    Zero-shot   & 0.45 & 0.79 & 0.52 & 0.50 & 0.67 & 0.53 & 86.2\% & 80.9\% & 0.000\% \\
    CoT         & 0.44 & 0.77 & 0.51 & 0.50 & 0.64 & 0.52 & 84.5\% & 79.2\% & 0.006\% \\
    \textbf{Few-shot-5}
                & \textbf{0.44} & \textbf{0.79} & \textbf{0.52}
                & \textbf{0.51} & \textbf{0.68} & \textbf{0.54}
                & \textbf{85.3\%} & \textbf{81.1\%} & \textbf{0.006\%} \\
    \bottomrule
  \end{tabular}
\end{table*}

\noindent\textbf{Evaluation on Benign PCAPs.}
On the external 191-PCAP benign benchmark, none of which contains a security-relevant flow, fingerprinting removes 98.0\% of flows and filters 126 PCAPs entirely, requiring no LLM call. Of the remaining 65, exactly one yields a rule, flagging a deprecated TLS~1.0 connection rather than malware behavior. \system{} reaches 99\% flow-classification accuracy and emits no malware-style rule across the 191 benign executions.

\noindent\textbf{Generalizability of rules}
A rule that fires on its source capture may be memorizing it. Replaying each
capture's rules against every other and labeling pairs same- or other-family,
\system{} reaches FAS-F1 0.67 on same-family targets against 0.75 for the
expert-authored rules, triggering slightly more often (0.58 versus 0.55). On
other-family targets both collapse and \system{} falls further (0.06 versus
0.16), so the extra triggers are not indiscriminate.  Rules therefore transfer to
unseen same-family samples at near-reference accuracy while retaining
cross-family specificity. Table~\ref{tab:family-generalization} summarizes these rates: on same-family targets, few-shot-5 triggers slightly more often than ground truth (0.58 vs.\ 0.55) but with lower precision (0.83 vs.\ 1.00) and recall (0.57 vs.\ 0.60), so generated rules alert on some flows that are not security-relevant while missing a portion of those that are. Ground-truth precision is 1.00 in both relations by construction: a target's security-relevant flows are defined as the flows ET~Open alerts on, so ground-truth rules cannot alert outside that set. On other-family targets, few-shot-5 triggers more often than ground truth (0.09 vs.\ 0.07), yet recall drops sharply (0.09 to 0.03), indicating the additional triggers predominantly fire on flows not security-relevant to the target family. A per-family-pair breakdown, including the heatmap of trigger-rate deviations (Figure~\ref{fig:generalizability}), is in Appendix~\ref{app:family-generalization}.

\begin{table}[t]
  \caption{Family generalization on the 1,074-source non-singleton subset.
  Metrics are micro-averaged over source--target pairs.}
  \label{tab:family-generalization}
  \small
  \centering
  \resizebox{\columnwidth}{!}{%
  \begin{tabular}{@{}llcccc@{}}
    \toprule
    \textbf{Relation} & \textbf{Rule Source} &
    \shortstack{\textbf{Trigger}\\\textbf{Rate}} &
    \shortstack{\textbf{FAS}\\\textbf{Precision}} &
    \shortstack{\textbf{FAS}\\\textbf{Recall}} &
    \shortstack{\textbf{FAS}\\\textbf{F1}} \\
    \midrule
    Same-family  & Ground Truth & 0.55 & 1.00 & 0.60 & 0.75 \\
    Same-family  & Few-shot-5   & 0.58 & 0.83 & 0.57 & 0.67 \\
    \midrule
    Other-family & Ground Truth & 0.07 & 1.00 & 0.09 & 0.16 \\
    Other-family & Few-shot-5   & 0.09 & 0.38 & 0.03 & 0.06 \\
    \bottomrule
  \end{tabular}
  }
\end{table}

\subsection{RQ2: What Determines Rule Quality}
\label{sec:rq2-determinants}

RQ1 establishes that the pipeline works, not why. Rule quality could
come from \system{}'s own components, from the general strategy of scaffolding
an LLM at all, or from the backbone model itself. 

We isolate these three explanations in turn: (i) an internal ablation removes \system{}'s own components (benign filtering, the verification agent, and the repair agent) one at a time to measure each component's contribution, and separately swaps the backbone under a fixed scaffold to measure how much the
backbone itself matters; (ii) a cross-system comparison holds the backbone fixed and swaps \system{}'s scaffold for two general-purpose coding agents, testing whether a task-specific scaffold is necessary or a general-purpose one suffices; and (iii) a comparison against a state-of-the-art model and scaffold
asks what the best available combination can achieve relative to ours.

\subsubsection{Isolating \system{}'s Own Components}
\label{sec:rq2-components}

We isolate the contribution of each component in turn to see how much they contribute in rule quality.

\noindent\textbf{Benign traffic filtering.}
\label{sec:results-fingerprinting}
\system{} removes background flows before the Rule Synthesis Agent sees them.
Across 1,296 PCAPs, the dataset contains 131,787 flows. Of these, 6,937
(5.3\%) are security-relevant, meaning they trigger at least one Suricata
rule. Benign fingerprinting removes 110,879 flows (88.8\%) while discarding
only 142 (2.0\%) of the security-relevant flows. This cuts the median flows
per PCAP from 65 to 12.

\begin{figure*}[t]
    \centering
    \includegraphics[width=\textwidth,height=0.32\textheight,keepaspectratio]{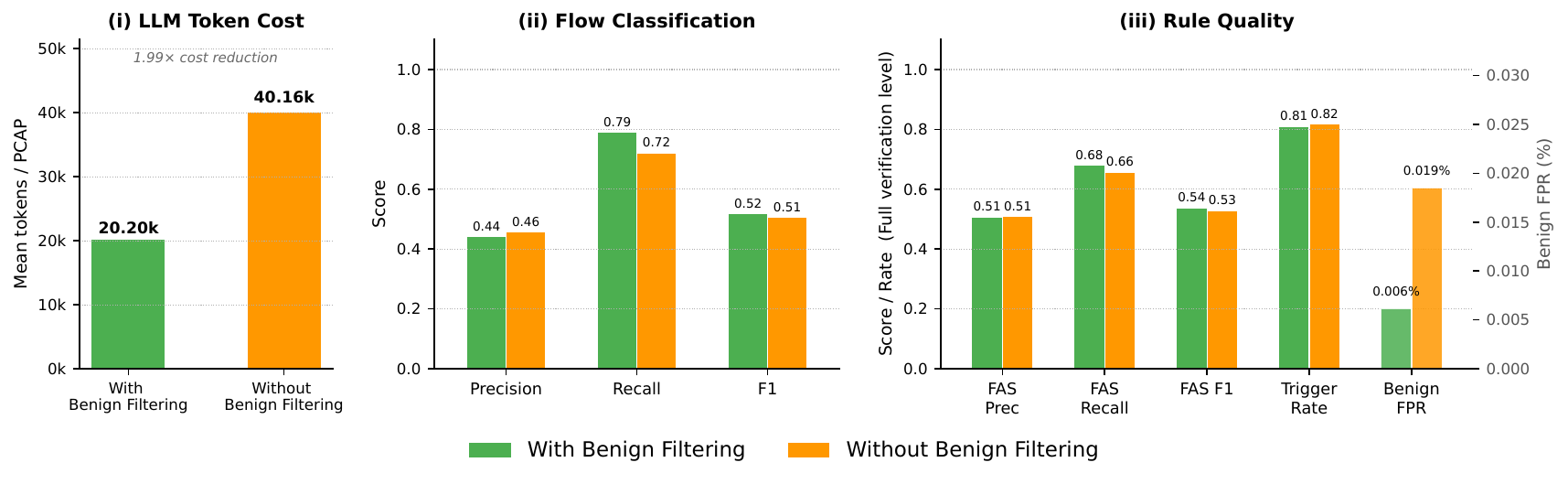}
    \caption{Impact of benign traffic filtering on (i) LLM token cost, (ii) flow-classification performance, and (iii) rule quality.}
    \label{fig:filtering-impact}
\end{figure*}

As Figure~\ref{fig:filtering-impact} shows, filtering halves cost without
hurting quality. Mean tokens per PCAP fall from 40.16k to 20.20k, a
1.99$\times$ reduction, and benign FPR falls from 0.019\% to 0.006\%. Flow
classification shifts along the precision--recall tradeoff rather than
improving outright. This happens because the unfiltered prompt warns that
many flows are routine background traffic, making the model more cautious
about flagging them. Fewer flags means higher precision (0.46) but lower
recall (0.72). Once background flows are removed, that warning is gone, so
the model flags more freely. Recall rises to 0.79 but precision falls to
0.44. The LLM classifies flows competently on its own. Filtering simply buys
the same overall quality (FAS-F1 0.53 to 0.54, F1 0.51 to 0.52) at half the
cost.

\noindent\textbf{Verification-agent ablation.}
We isolate the contribution of each stage in the verification pipeline by
holding the synthesis output fixed and enabling stages one at a time.
Table~\ref{tab:ablation-fs5} shows that each stage catches a different kind
of bad rule. Syntax validation catches rules Suricata cannot parse, raising
the pass rate from 81.8\% to 85.2\%. Since these rules were unusable
regardless of detection quality, removing them barely moves FAS-F1. Malware
replay tests each rule against the source malware traffic and discards rules
that never fire, so it removes rules that pass syntax but detect nothing.
This is what drives detection quality, raising FAS-F1 from 0.456 to 0.537
through gains in both precision and recall. Benign replay tests each rule
against benign traffic and discards rules that misfire on it. Since these
rules already detect malware correctly, removing them barely changes FAS-F1,
but it cuts benign FPR from 0.128\% to 0.006\%. Together, the three stages
take FAS-F1 from 0.443 to 0.539.

\begin{table}[t]
\centering
\caption{Verification-agent ablation on 1,296 PCAPs. Stages are enabled cumulatively.}
\label{tab:ablation-fs5}
\small
\setlength{\tabcolsep}{3pt}
\begin{tabular*}{\columnwidth}{@{\extracolsep{\fill}}lcccccc@{}}
\toprule
& \multicolumn{3}{c}{FAS} & \multicolumn{3}{c}{Rule Quality} \\
\cmidrule(lr){2-4} \cmidrule(l){5-7}
Stage & P & R & F1 & Syn. & Trig. & FPR \\
\midrule
None            & .435 & .531 & .443 & 81.8\% & 72.1\% & 0.091\% \\
+\,Syntax       & .445 & .549 & .456 & 85.2\% & 71.5\% & 0.091\% \\
+\,Malware      & .506 & .679 & .537 & 85.2\% & 80.9\% & 0.128\% \\
+\,Benign       & .507 & .680 & .539 & 85.3\% & 81.1\% & 0.006\% \\
\bottomrule
\end{tabular*}
\end{table}

\noindent\textbf{Repair agent effectiveness.}
\label{sec:repair-agent}
Across the corpus, 2,530 rules are generated and 2,091 (82.6\%) pass every check
immediately. The repair agent attempts 247 fixes and succeeds on 124, a 50.2\%
success rate, with malware-trigger repairs landing more often (61\%) than syntax
repairs (40\%). Two residues remain: 37 rules hit Suricata errors with no
knowledge-base entry and failed under generic prompting, and 25 had their syntax
fixed by keyword substitution while their content predicates stayed on the wrong
sticky buffer, so replay still failed.
Table~\ref{tab:repair-kb} breaks outcomes down by failure type.

\begin{figure*}[t]
    \centering
    \includegraphics[width=\textwidth]{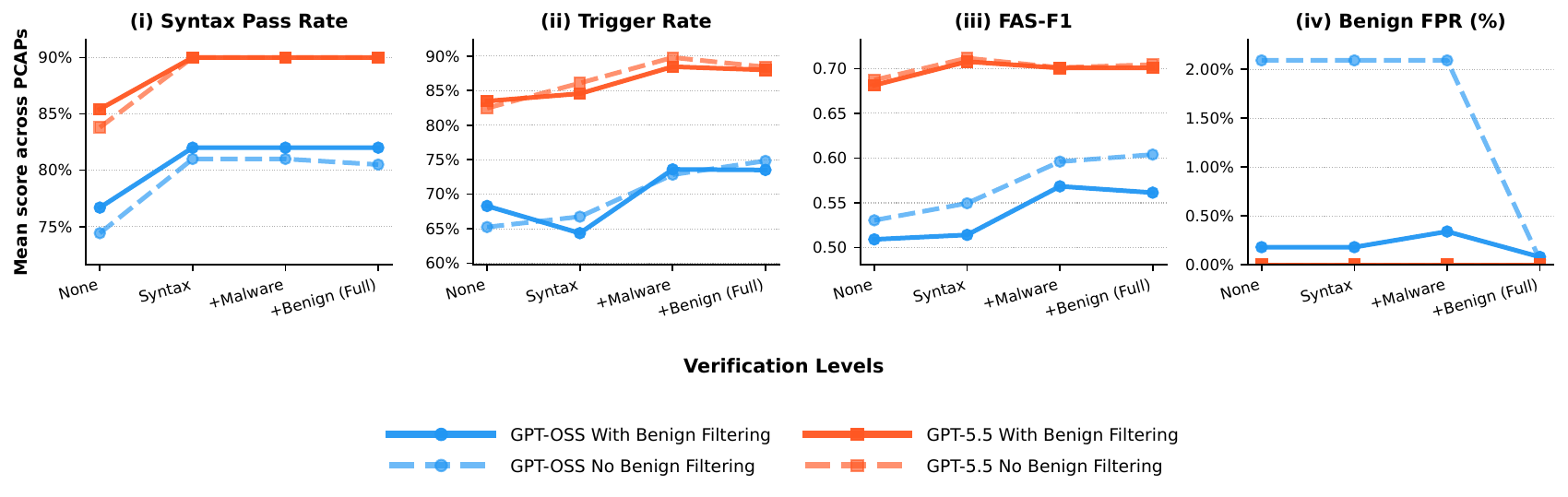}
     \caption{Cumulative verification progression for GPT-5.5 and GPT-OSS-120b under few-shot-5 prompting, with and without benign-flow filtering. The x-axis shows enabled stages: \emph{None} uses raw rules; \emph{Syntax} adds Suricata syntax validation and repair; \emph{+Malware} adds replay on the source malware PCAP; and \emph{+Benign (Full)} adds benign replay to suppress false positives.}
    \label{fig:gpt5-gptoss-100sha}
\end{figure*}

\noindent\textbf{Backbone sensitivity within \system{}.}
We test whether swapping \texttt{gpt-oss-120b} for a frontier model changes
how \system{} performs. The ablations above hold the backbone fixed and vary
the scaffold; this test does the opposite, holding the scaffold fixed and
swapping the backbone for GPT-5.5 (Figure~\ref{fig:gpt5-gptoss-100sha}), on a
family-stratified 100-PCAP subset sampling up to three PCAPs per malware
family until reaching 100.
GPT-5.5 outperforms \texttt{gpt-oss-120b} across every metric we track,
most notably raising FAS-F1 from 0.561 to 0.701, so a stronger backbone
does raise the pipeline's ceiling.
What stays the same is what each verification stage is for. Syntax checking
still does nearly all of the work of getting rules to parse, moving the
syntax pass rate from 76.7\% to 82.0\% for \texttt{gpt-oss-120b} and from
85.4\% to 90.0\% for GPT-5.5, before either model's rules are ever tested
against traffic. Malware replay still drives whether a rule actually fires
on the malware it was written for, lifting the trigger rate for both models
by a similar margin. Benign replay still does the work of suppressing false
positives, taking \texttt{gpt-oss-120b}'s unfiltered benign FPR from 2.09\%
down to 0.05\% and keeping GPT-5.5 close to zero throughout. In short, the
backbone changes how high \system{} can reach, but not what each stage of
the pipeline contributes to getting there.

\subsubsection{Our Scaffold vs. Other Scaffolds, Same Backbone}
\label{sec:rq2-scaffolds}
Rule quality could come from \system{}'s task-specific design, or simply
from wrapping any capable agent around the same backbone model. To tell
these apart, we hold the backbone fixed at \texttt{gpt-oss-120b} and compare
\system{} against two general-purpose coding agents, OpenClaw and Hermes
Agent, given the same task and prompt. Table~\ref{tab:arms} reports all six systems. The other three, Claude
Code, the naive regex extractor, and \system{} on the Claude Opus~5
backbone, are analyzed in \S\ref{sec:rq2-backbone}. This section isolates
the three systems sharing the \texttt{gpt-oss-120b} backbone.

\begin{table*}[t]
\centering\small
\caption{Cross-system comparison on 200 malware PCAPs. \emph{Cover} is the fraction of
captures for which the system emits at least one rule. \emph{Over-fire} is
alerted flows divided by ground-truth flows.}
\label{tab:arms}
\begin{tabular}{@{}lrrrrrrr@{}}
\toprule
\textbf{System (backbone)} & \textbf{Cover} & \textbf{FAS-F1}
& \textbf{FAS-P} & \textbf{FAS-R} & \textbf{Rules/PCAP}
& \textbf{Over-fire} & \textbf{Tokens/capture} \\
\midrule
\textbf{\system{}} (Claude Opus~5) &  95\% & \textbf{0.656} & \textbf{0.618} & 0.830 & 3 & 1.49$\times$ & 30{,}910 \\
Claude Code (Claude Opus 5)     & 100\% & 0.623 & 0.536 & 0.967 & 3 & 10.99$\times$ & 1{,}245{,}898 \\
\textbf{\system{}} (gpt-oss-120b) &  79\% & 0.524 & 0.523 & 0.610 & 2 & 1.17$\times$ & 23{,}836 \\
OpenClaw (gpt-oss-120b)         & 100\% & 0.416 & 0.379 & 0.575 & 4 & 2.07$\times$ & 885{,}734 \\
Hermes Agent (gpt-oss-120b)     &  98\% & 0.400 & 0.355 & 0.575 & 4 & 2.50$\times$ & 1{,}192{,}519 \\
Regex, naive (none)         & 100\% & 0.195 & 0.130 & 0.809 & 31 & 8.18$\times$ & 0 \\
\bottomrule
\end{tabular}
\end{table*}

\noindent\textbf{General-purpose scaffolds are interchangeable; a task-specific
one is not.} OpenClaw and Hermes are different agents on the same model and
prompt, and they land in the same place: FAS-F1 0.416 versus 0.400, with recall
identical at 0.575. Both write rules that stay generic. Neither uses \texttt{bsize}, which pins a matched field to an exact byte
length. Neither attaches \texttt{fast\_pattern}, which flags the most
distinctive byte string in a rule, to more than a fifth of their rules.
Both keywords are available to every system we evaluate. Their absence
reflects a failure to identify distinguishing patterns, not a lack of
access. \system{}, at the same backbone, reaches 0.524 FAS-F1 with
\texttt{fast\_pattern} on 48\% of rules. That is ahead of the two agents
by 0.109 and 0.125. \system{} also emits half as many rules per capture.
It alerts on 1.17$\times$ the reference flow count, against
2.1--2.5$\times$ for OpenClaw and Hermes.

\noindent\textbf{Abstention and cost.}
\system{} emits no rules for 42 of 200 captures, 34 of them because the
synthesis agent classified every flow as background. When \system{}
abstains, that capture contributes zero precision and zero recall to the
average, since no rule was emitted to be scored. This pulls down
\system{}'s system-wide precision and recall relative to what it achieves
when it does commit to an answer. Averaged across all 200 captures,
precision is 0.523. Restricted to the captures where \system{} emits at
least one rule, precision rises to 0.662, the highest among the three
same-backbone systems. Abstention concentrates in the stateful stratum,
where evidence must be aggregated across many flows rather than read from
one, and \system{} declines rather than emit a weakly supported rule. This
selectivity is also cheap. \system{} processes 23{,}836 tokens per capture
against 885{,}734 for OpenClaw on the same open model, a 37$\times$
reduction, for 0.144 more precision. Selecting flows before generation and
verifying rules after it is, on this corpus, both cheaper and more precise
than letting a general-purpose agent explore the capture unguided.

\subsubsection{Backbone and Baseline: From Regex to a Frontier Model}
\label{sec:rq2-backbone}

The comparison above holds the backbone fixed to isolate scaffold effects.
To see what lies at either end of that comparison, we bracket it with two
systems outside it: a naive regex extractor with no model at all, showing
what selection-free coverage looks like, and Claude Code, a general-purpose
coding scaffold paired with a frontier model instead of a weak one, showing
what a stronger backbone can add. We then swap scaffolds at that same
frontier backbone, pairing our own scaffold with Claude Opus~5 as well.
Table~\ref{tab:arms} reports all three of these systems alongside the three
same-backbone systems above.

\noindent\textbf{No intelligence, no selection.} The naive regex extractor
emits one rule per protocol-field indicator it observes anywhere in the
capture, with no judgment about whether that indicator is malicious. It
reaches the highest recall of any system, 0.809, simply by proposing a rule
for everything, but its precision collapses to 0.130 and it alerts on
8.18$\times$ the reference flow count, 31 rules per capture on average.
FAS-F1 lands at 0.195, the lowest of any system: without selection, coverage
is worthless, since the reference ruleset's own selectivity is exactly what
the metric rewards.

\noindent\textbf{A frontier model raises the ceiling but is not required
for the precision.} Claude Code pairs Claude Opus~5 with the same kind of
general-purpose scaffold that stalled around FAS-F1 0.40 on
\texttt{gpt-oss-120b} for OpenClaw and Hermes (\S\ref{sec:rq2-scaffolds}).
On the stronger model, that same scaffold class reaches 0.623, the best
of any system. That lead comes almost entirely from recall. It answers
every capture at 0.967 recall, against \system{}'s 0.610. Precision tells
a different story. Claude Code's 0.536 and \system{}'s 0.523 are
essentially tied, but they get there differently. Claude Code alerts on 10.99$\times$ the reference flow count to reach
that precision, while \system{} alerts on only 1.17$\times$. This is a different operating point, not a better one: ten alerts per
real flow is the kind of ratio that drives alert fatigue.

\noindent\textbf{\system{}'s scaffold at the frontier backbone.} We can hold the backbone fixed and swap scaffolds directly, pairing
\system{}'s own scaffold with the same Claude Opus~5 backbone that powers
Claude Code. At equal backbone, \system{} reaches FAS-F1 0.656, ahead of Claude Code's
0.623, with FAS-precision 0.618 against 0.536 and an over-fire ratio of
1.49$\times$ against 10.99$\times$. \system{} trades some recall for this
gain, 0.830 versus 0.967, and answers 95\% of captures against Claude
Code's 100\%. Our scaffold declines rather than emit a weakly supported
rule. Because the backbone is now identical, this result isolates the scaffold
as the source of the gap. \textit{A task-specific scaffold does not just close
the distance to a frontier general-purpose agent; it surpasses it, at a
40$\times$ reduction in tokens per capture}.

\noindent\textbf{Why the cost gap matters.} \system{} runs on an
open-weight backbone that an organization with sufficient compute can
self-host, avoiding both per-token billing and the need to send network
traffic to a third-party API. Claude Code's cost and privacy profile are
tied to a closed, API-served model. The token gap above is therefore not
just a cost difference; it reflects whether sensitive traffic ever has to
leave the organization.

\section{RQ3: Rule Generation Under Encryption}
\label{sec:encryption}

This section asks how \system{} performs once traffic is encrypted.
Encryption hides the payload, but leaves some handshake fields visible.
We measure what \system{} can still do with what remains, and how well
the rules it writes hold up.

\noindent\textbf{Performance across the encryption spectrum.}
We classify each capture by the share of its flow records that Suricata
dissects as \texttt{tls} or \texttt{quic}, the two encrypted protocols
with enough volume in this corpus to support separate analysis. This
classification is based on protocol dissection rather than which port
traffic uses, so encrypted traffic on a non-standard port is still
correctly identified as encrypted. Based on this share, we sort captures into three bands: cleartext (mostly
unencrypted, 331 captures), mixed (a blend of encrypted and unencrypted
traffic, 889 captures), and encrypted (mostly encrypted, 76 captures).
The bands differ mainly in how much traffic they hide, not in which
protocols show up: nearly all mixed and encrypted captures still contain
both TLS and HTTP, and every capture in the corpus contains DNS. Table~\ref{tab:encryption-bands} reports \system{}'s performance in each
band. As encryption increases, FAS-F1 drops from 0.614 on cleartext to
0.355 on encrypted, and precision drops faster than recall. In other
words, rules written under encryption still catch a similar share of the
ground truth, but they also raise more false alerts along the way.
Coverage moves in the opposite direction, rising from 80.4\% to 88.2\%,
and the rules the pipeline does write fire more reliably as encryption
increases. \system{} does not go quiet as payload disappears; it answers
more often and more narrowly, at some cost to precision.

\begin{table}[t]
\centering\footnotesize
\caption{\system{} performance by encryption band across the full
1,296-PCAP corpus (few-shot-5, full verification). Bands are the share of
Suricata-dissected flow records carrying \texttt{tls} or \texttt{quic}.
\emph{Cover} is the fraction of captures for which at least one rule
survives verification, and \emph{Trig.} is the trigger rate over those
captures.}
\label{tab:encryption-bands}
\setlength{\tabcolsep}{3pt}
\begin{tabular*}{\columnwidth}{@{\extracolsep{\fill}}lrrrrrr@{}}
\toprule
Band & PCAPs & Cover & FAS-P & FAS-R & FAS-F1 & Trig. \\
\midrule
Cleartext & 331 & 80.4\% & 0.603 & 0.677 & 0.614 & 91.1\% \\
Mixed     & 889 & 86.8\% & 0.485 & 0.701 & 0.526 & 96.3\% \\
Encrypted &  76 & 88.2\% & 0.349 & 0.452 & 0.355 & 96.9\% \\
\midrule
All       & 1{,}296 & 85.3\% & 0.507 & 0.680 & 0.539 & 95.1\% \\
\bottomrule
\end{tabular*}
\end{table}

\noindent\textbf{What \system{} writes when the payload is encrypted.}
Of the 40 samples whose ground truth requires a content match on a fully
encrypted trace, listed as \emph{Content, encrypted} in
Table~\ref{tab:strata200}, \system{} answers 34 and writes 89 rules. Of those,
26 match on \texttt{tls.sni}, the name the session announces in its handshake,
and 38 on \texttt{dns.query}, the name it resolves beforehand. Stateful rules
are built on the same names: 13 of the 27 written on these samples combine
their threshold with a \texttt{tls.sni} match, against 5 of 28 on the cleartext
samples. Both forms of detection therefore remain available once the payload is
gone, content rules written against the handshake and DNS fields that stay
visible, and stateful rules written against behavior that never needed the
payload at all.

\section{Discussion}
\label{sec:discussion}

\noindent\textbf{Scaffolding is what makes local deployment viable.}
The choice to run an open-weight model in-house is not just a cost
decision. It is often the only option when the traffic itself is the
constraint. Malware captures, and the network layout they reveal, can be
sensitive operational information an organization cannot send to a
third-party API regardless of that API's quality. Under that constraint,
the question this paper answers changes from ``which model is best'' to
``how much of a frontier model's advantage can a well-designed scaffold
recover on a model you are allowed to run.'' On the open-weight backbone,
our scaffold comes within 0.099 F1 of the frontier system at a fraction
of the cost. At equal backbone, it surpasses that system outright,
showing the scaffold is not just recovering a frontier model's advantage
but is itself the larger source of it.

\noindent\textbf{What happens when more fields are encrypted.}
Every result in this section rests on a name: the server name in the TLS
handshake or the DNS query that precedes it. Wider adoption of Encrypted
Client Hello~\cite{ech} and DNS over HTTPS or TLS~\cite{rfc8484,rfc7858}
would remove both from what a sensor can observe, leaving \system{}'s
content and rate rules without the anchor they currently key on. In that
setting, rule generation would need to lean more heavily on stateful rules
that read overall network behavior, such as connection timing and volume,
rather than any single field in the traffic. 

\noindent\textbf{Where FAS goes to zero, and why.}
Of 1{,}296 PCAPs, 306 score FAS-F1 of zero: 211 produce no valid rule and 95
produce valid rules that miss every security-relevant flow. Two structural
causes account for them.
\newline\textbf{(i) Externally attributed indicators.} Signatures keyed
on JA3/JA3S fingerprints expose no interpretable pattern in the traffic.
Without a live intelligence feed, no generator can decide whether an
unseen fingerprint is malicious. The limitation is missing attribution,
not rule generation.
\newline\textbf{(ii) Over-generation.} Of 1,296 PCAPs, a further 207
PCAPs land in the partial-alignment band ($0 < \text{FAS-F1} < 0.5$), at
mean precision $0.28$ and recall $0.77$. Rather than one rule aligned
with the target, the model emits several for different protocol signals
in the same capture. Recall survives when one of these rules matches, but
precision suffers when the others hit background flows.

\noindent\textbf{What FAS does not measure.}
FAS scores a generated rule by its flow-level agreement with the ET~Open
rule that fired on the same capture, treating that rule as ground truth
rather than as one valid detection among others. A generated rule that matches on a different but equally valid indicator
scores as a false positive, and one that would detect the family more
robustly than the expert rule cannot score above it. Agreement alone also cannot separate a rule that
fires twice too often from one that fires fifty times too often, which is
why we report over-fire alongside FAS. Whether agreement with a curated
ruleset tracks operational detection quality is a question this corpus
cannot answer without alert data from a live deployment.

\section{Related Work}
\label{sec:related}

We organize related work into three categories: traditional automated signature generation, LLM-based rule generation systems, and LLMs for network traffic analysis.

\subsection{Traditional Signature Generation}

Honeycomb~\cite{kreibich2004honeycomb} extracts signatures from honeypot traffic by longest common substring, Autograph~\cite{kim2004autograph} clusters gateway payloads for worm signatures, and Polygraph~\cite{newsome2005polygraph} resists polymorphism through conjunctions of invariant substrings. These showed that malicious traffic carries reusable invariants, but they emit byte-level signatures rather than Suricata rules with protocol-aware keywords, flow constraints, and metadata; more recent deep-learning IDS work~\cite{xu2025dlids} outputs labels or anomaly scores rather than rule artifacts. Our non-LLM baselines are a modern analogue: they extract every observable protocol-field indicator and emit a rule for each. That they stay competitive on cleartext and degrade sharply under encryption (\S\ref{sec:encryption}) suggests the historical difficulty was invariant \emph{extraction}, whereas under encryption it is invariant \emph{selection}.

\subsection{LLM-Based Rule Generation Systems}

\noindent\textbf{RuleMaster+}~\cite{rulemaster2025} fine-tunes an LLM to generate
IDS rules from proof-of-concept exploit descriptions, showing that LLMs learn
rule-writing conventions but assuming curated exploit text as input.

\noindent\textbf{FALCON}~\cite{falcon2025} generates Snort and YARA rules from
CTI reports using an agentic workflow with explicit validation. Both start from
structured intelligence rather than traffic, so neither addresses the challenge
that dominates our setting: identifying the security-relevant flows inside a
noisy capture before any rule can be written.

\noindent\textbf{Moreno et al.}~\cite{moreno2025leveraging} are the closest
PCAP-based prior work, generating Suricata rules from captures of controlled
industrial attack scenarios under zero-shot, few-shot, and chain-of-thought
prompting. Their setting isolates anomalous flows by attacker IP address, so the
LLM receives traffic already filtered toward the attack, whereas \system{} starts
from noisy sandbox PCAPs where identifying the relevant flows is part of the
task. Their repair loop is parser-error driven; ours adds execution feedback from
malware replay and benign testing, which is worth $+0.083$ FAS-F1 over
syntax-only verification in our ablation (0.456 to 0.539).

\noindent\textbf{GRIDAI}~\cite{gridai2025} is a multi-agent framework for
ruleset \emph{evolution}, deciding whether an incoming Web-attack sample is a new
type or a variant already covered and then generating or repairing accordingly,
evaluated on HTTP attack samples. \system{} instead targets discovery from noisy
multi-protocol sandbox traces, where the relevant flows must be identified before
any rule is written.

\noindent\textbf{Other recent systems.} RuleXploit~\cite{rulexploit2025}, Hex2Sign~\cite{hex2sign2024}, GenTI~\cite{genti2026}, and \cite{unirule2026} generate rules from exploit descriptions, hexadecimal payloads, and unseen-attack benchmarks, scoring with BERTScore or an LLM judge in place of execution. The distinction that matters here is whether a rule is ever run: model-judged similarity to a reference does not establish that a rule loads, fires on its source traffic, or stays quiet on benign traffic. Consistent with this, \cite{beyondsyntax2026} finds practitioners treat execution evidence as a precondition for adoption.

\noindent\textbf{General-purpose agents.} Agents that operate over a filesystem and shell appear in our evaluation as baselines rather than related systems. They are not built for rule generation and receive our prompt rather than their own. Their role is methodological, establishing what a capable model does with the same instructions and evidence but no task-specific scaffold. To our knowledge no prior work on IDS rule generation holds the backbone fixed while varying the surrounding system, so the scaffold's contribution has not previously been separated from the model's.

\subsection{LLMs for Network Traffic Analysis}
Recent work applies LLM-style models to network traffic understanding.
TrafficLLM~\cite{trafficllm2024} adapts open-source LLMs to raw traffic data,
while NetGPT~\cite{netgpt2023} pretrains generative models over packet- and
flow-level representations. These systems show that language-model-style
architectures can learn useful traffic representations. However, their goals are
traffic analysis and synthesis, not deployable IDS rule generation. \system{}
focuses specifically on transforming raw network PCAPs into validated Suricata signatures.

\section{Conclusion}
\label{sec:conclusion}

We presented \system{}, an agentic framework that synthesizes deployable
Suricata rules directly from network traffic. We used it to ask what
determines rule quality: the backbone model or the scaffold around it. Evaluated across 1,296 malware PCAPs, execution-grounded verification and
benign traffic fingerprinting together raise rule quality while cutting
both cost and false positives, showing the pipeline scales to a large,
real-world corpus. Under encryption, precision falls as payload
disappears, and generation leans more on stateful rules and header-level
fields such as the TLS SNI and the DNS query that precedes a connection.
Encryption has not eliminated the need for content-based rules, though:
ET~Open itself continues to publish plaintext-targeting signatures today,
a sign that a meaningful share of malware still operates over unencrypted
traffic. The central finding is about the scaffold, not the backbone. An open, self-hosted model paired with a task-specific scaffold approaches frontier-model quality in generating deployable Suricata rules directly from network traffic, without sending sensitive traffic to a third-party API. This finding demonstrates that high-quality rule generation is possible without relying on a closed model. Using the same backbone, Claude Opus~5, \system{} outperforms Claude Code while consuming 40$\times$ fewer tokens per capture. Together, these results show that \system{} supports both private, self-hosted deployment and improved detection quality with frontier models, while substantially reducing token consumption and associated costs.

\noindent\textbf{Future Work.}
One natural extension is integrating live threat intelligence to cover
the one class of rule our pipeline cannot write on its own: indicators
with no interpretable pattern in the traffic itself, such as a JA3,
JA3S, or JA4 TLS fingerprint. A fingerprint is just a hash, so no amount of
reasoning over the capture can tell us whether an unseen one is malicious. Pairing the synthesis agent with an external feed, such as a
fingerprint or hash reputation database built from past incidents, would
supply that missing evidence. The model would then focus on deciding whether a matched fingerprint,
combined with the rest of the capture's behavior, is enough reason to
write a rule.

\bibliographystyle{plain}
\bibliography{references_url}

\appendix
\section*{Ethical Considerations}
\label{app:ethics}

\noindent\textbf{Malware samples.}
Malware artifacts used in this study were sourced from MalwareBazaar~\cite{malwarebazaar}, ANY.RUN, and Triage. Where binaries were collected (MalwareBazaar and ANY.RUN), samples were executed solely within isolated Cuckoo Sandbox~\cite{cuckoo} environments with no network connectivity to production systems; Triage-provided PCAPs were ingested directly as traffic traces. No malware binary files were distributed, deployed, or tested outside the sandboxed environment.

\noindent\textbf{Responsible use.}
While \system{} could in principle reveal information about malware detection capabilities (e.g., which network patterns are being monitored), the generated rules are standard Suricata signatures comparable to those already publicly available in the ET~Open ruleset.
We do not believe this work introduces novel dual-use risks beyond those inherent to existing public rulesets.

\section*{Open Science}
\label{app:openscience}

\begin{enumerate}
    \item \textbf{Source code:} \url{https://anonymous.4open.science/r/rulepilot-artifact-DB05}. Includes the \system{} pipeline, evaluation scripts, and the baseline harness used for the cross-system.
    \item \textbf{Dataset.} We do not release the malware corpus or binaries for the reasons described in the Ethical Considerations section.
\end{enumerate}

\section{Dataset and Ground-Truth Details}
\label{app:dataset-details}

This appendix provides additional details on the malware corpus, the ET~Open
rules used to construct ground-truth labels, and the normalized malware-family
distribution used in our evaluation.

\subsection{Temporal Coverage of Malware PCAPs}
\label{app:malware-temporal-coverage}

We report the temporal coverage of the malware corpus to show that the benchmark
includes both older and recently submitted samples. Table~\ref{tab:malware-temporal-distribution} groups malware PCAPs by VirusTotal \texttt{first\_submission\_date}.

\begin{table}[t]
\centering
\caption{Temporal coverage of malware PCAPs. Malware PCAPs are grouped by VirusTotal \texttt{first\_submission\_date}.}
\label{tab:malware-temporal-distribution}
\footnotesize
\setlength{\tabcolsep}{6pt}
\renewcommand{\arraystretch}{1.05}
\begin{tabular}{@{}lr@{}}
\toprule
\textbf{Period} & \textbf{Malware PCAPs} \\
\midrule
2014--2015 & 7   \\
2016--2017 & 64  \\
2018--2019 & 32  \\
2020--2021 & 44  \\
2022--2023 & 24  \\
2024--2025 & 126 \\
2026 & 999 \\
\midrule
\textbf{Total} & \textbf{1,296} \\
\bottomrule
\end{tabular}
\end{table}

\subsection{Temporal Coverage of Triggered ET~Open Rules}
\label{app:rule-temporal-coverage}

We also examine the age of the ground-truth ET~Open signatures triggered by the
malware corpus. Table~\ref{tab:rule-temporal-distribution} groups unique
triggered ET~Open rules by their \texttt{created\_at} metadata field.

\begin{table}[t]
\centering
\caption{Temporal coverage of unique triggered ET~Open rules. Rules are grouped by the ET~Open \texttt{created\_at} metadata field.}
\label{tab:rule-temporal-distribution}
\footnotesize
\setlength{\tabcolsep}{6pt}
\renewcommand{\arraystretch}{1.05}
\begin{tabular}{@{}lr@{}}
\toprule
\textbf{Period} & \textbf{Unique Triggered Rules} \\
\midrule
2010--2011 & 21  \\
2012--2013 & 22  \\
2014--2015 & 28  \\
2016--2017 & 26  \\
2018--2019 & 53  \\
2020--2021 & 42  \\
2022--2023 & 72  \\
2024--2025 & 336 \\
2026 & 17  \\
\midrule
\textbf{Total} & \textbf{617} \\
\bottomrule
\end{tabular}
\end{table}

\subsection{Malware Family Distribution}
\label{app:malware-family-distribution}

To characterize family diversity, we group normalized malware-family labels by
the number of PCAPs assigned to each family. Table~\ref{tab:malware-family-size-distribution}
summarizes this distribution and reports PCAPs without an assigned family label as \textsc{SINGLETON} instances.

\begin{table*}[t]
\centering
\caption{Distribution of normalized malware-family sizes. PCAPs without an assigned family label are treated as \textsc{SINGLETON} instances and reported separately. Family names are grouped by the number of PCAPs assigned to each family.}
\label{tab:malware-family-size-distribution}
\scriptsize
\setlength{\tabcolsep}{4pt}
\renewcommand{\arraystretch}{1.08}
\begin{tabular}{@{}lrrp{0.68\textwidth}@{}}
\toprule
\textbf{Samples / family} & \textbf{\# Families} & \textbf{\# PCAPs} & \textbf{Families} \\
\midrule
100+ & 1 & 271 &
asyncrat \\

41--100 & 6 & 341 &
vidar, remcos, sfuzuan, xworm, corewarrior, futu \\

11--40 & 12 & 181 &
autoinject, wacatac, cryptnot, agenttesla, salat, quasar, blank, connectwise, lummastealer, virlock, gandcrab, stealc \\

3--10 & 64 & 335 &
dorshel, formbook, salatstealer, minix, bookworm, heracles, shifu, lolbot, zorex, sagent, teslacrypt, disco, maskgramstealer, bumblebee, crypt0l0cker, birddog, netwire, socgholish, njrat, purelogs, locky, upatre, babar, taskun, atbot, wannacry, buterat, redline, nuitka, havoc, bladabindinet, shade, ravartar, blankgrabber, houdini, acsogenixx, lummac, dbadur, vnfraye, disfa, egairtigado, fragtor, metastealer, amadey, kasidet, fabookie, xmrig, mikey, cobaltstrike, nanocore, qwexlafiba, lodarat, trickbot, azorult, raccoon, venomrat, netsupport, nemesis, raccoonstealer, morphine, black, emotet, alevaul, wanker \\

1--2 & 109 & 153 &
allaple, telebot, injuke, cryptowall, midie, injectornett, snakekeylogger, hlux, multiplug, znyonm, lethic, nemucod, ctblocker, warezov, cleanuploader, mydoom, darkcrystal, boxter, fugrafa, ligooc, strab, guloader, cerbu, oyster, gootkit, blind, hangover, jalapeno, zmutzy, obfdldr, hijackloader, socstealer, valleyrat, neutrino, lummac2, clipbanker, artifact, smokeloader, vobfus, ammyy, darkcomet, downeks, kelios, discordbot, windigo, denes, inoci, mars, amatera, discordrat, hiddentear, neshta, sdrop, risepro, backdoornjrat, brresmon, kepavll, blamon, makoob, lumma, cozyduke, barys, lockfile, privateloader, dridex, mofksys, crysan, tinynuke, sphinx, umbralstealer, bobik, stelpak, crypren, alvaro, aurastealer, spymax, mami, icedid, danabot, koobface, zegost, zombie, khalesi, dapato, medusalocker, obfsobjdat, bazloader, rhadamanthys, fatbeehive, sectoprat, nitol, kazuar, bladabindidldr, titanstealer, andromeda, banload, strelastealer, darkgate, cryxos, amnesia, polazert, expiro, lokibot, solmyr, ghostrat, multiverze, qqrob, hpqakbot, gcleaner \\
\midrule
\textbf{Subtotal: labeled families} & \textbf{192} & \textbf{1,281} & -- \\
\textsc{SINGLETON} / no family & -- & 15 & -- \\
\midrule
\textbf{Total} & \textbf{192} & \textbf{1,296} & -- \\
\bottomrule
\end{tabular}
\end{table*}

\section{Experimental Setup Details}
\label{app:setup-details}

\noindent\textbf{Model.}
The primary backbone is \texttt{gpt-oss-120b}, served through an
OpenAI-compatible institutional endpoint. We use an open model because network
flows may carry sensitive operational information that cannot be sent to a
third-party API, and because proprietary models become costly at corpus scale. We
additionally evaluate GPT-5.5 on a stratified 100-PCAP subset
(\S\ref{sec:rq2-determinants}). All runs
use temperature~0, a 65,536-token completion budget, and up to three repair
iterations per rule.

\noindent\textbf{Prompting strategies.}
We evaluate six strategies under identical generation parameters: zero-shot, chain-of-thought (CoT), few-shot with 1, 3, or 5 examples, and CoT+FS-3. Few-shot examples are (flow, rule) pairs drawn from real dataset PCAPs with their ET~Open rules across HTTP, DNS, TLS, TCP, and SMB, plus benign boundary examples carrying no rule; $N$ denotes malicious examples per protocol. Source PCAPs of the examples are excluded from evaluation. Templates are in Appendix~\ref{app:prompts}.

\noindent\textbf{Benign corpus.}
The 981 DIKE-derived benign PCAPs build the fingerprint library; a 600-PCAP subset serves benign replay during verification and a disjoint 200-PCAP subset final false-positive testing. An external 191-PCAP benchmark~\cite{Lucas21Malware}, used for none of these, evaluates benign-only abstention.

\noindent\textbf{Comparison systems.}
For RQ2 we run four additional systems over the 200-capture subset of
\S\ref{sec:strata}. Two are general-purpose coding agents driving the same
\texttt{gpt-oss-120b} backbone as \system{}: OpenClaw and Hermes Agent. One is
Claude Opus~5 under Claude Code. One is a non-LLM field extractor, 270 lines of
Python over \texttt{tshark} output, emitting one rule per indicator observed
anywhere in the capture with no benign filtering or model judgment. The three baseline agents receive
a byte-identical prompt, verified to one MD5 across all 603 sandbox copies
(Appendix~\ref{app:harness-prompt}), the same task description, and a script
that replays Suricata against its own capture only, in its own sandbox tree with
the capture hardlinked across trees so all five systems provably read the same
bytes.
That prompt states the task, the evidence constraint, and how rules are scored,
but it is not \system{}'s own pipeline prompt, which cannot be shared: the
pipeline prompt carries 31 few-shot examples quoting real ET~Open rules, so
handing it to a baseline would leak the ground truth. \S\ref{sec:results-known}
bounds the residual asymmetry, where moving from zero-shot to few-shot-5 is worth
0.018 FAS-F1 against the 0.109--0.125 gaps measured here. The design isolates one
variable at a time: OpenClaw versus Hermes swaps one general-purpose scaffold for
another at a fixed model and prompt, the open-model agents versus Claude Code
vary the model, and \system{} versus OpenClaw and Hermes replaces a
general-purpose scaffold with a task-specific one at a fixed model.

\noindent\textbf{Uniform benign false-positive control.}
\system{} discards every candidate rule that fires on a 600-PCAP benign corpus,
and the baselines have no such stage, which would confound any precision
comparison. We therefore apply the identical stage to every system after the
fact, replaying each system's emitted rules against the same 600 benign PCAPs
and re-scoring what survives. All RQ2 results are reported under this control, with
the uncontrolled scores alongside because the difference is itself a result.

\noindent\textbf{Token accounting.}
All \texttt{gpt-oss-120b} systems are measured at the API level through a
logging proxy with no prompt caching. Claude Code is measured from client transcripts and
does use caching, so we report both tokens processed and a billed-equivalent
figure charging cache reads at 10\% of input rate. Comparisons needing no such
correction, such as \system{} against OpenClaw on the same model, are noted where
they occur.

\noindent\textbf{Suricata configuration.}
All rule validation uses Suricata~8.0.3 on a local server equipped with
dual AMD~EPYC~7313 processors (32~physical cores, 64~logical~CPUs,
up to 3.7~GHz) and 1~TiB RAM.

\section{Flow Representation and Benign Fingerprinting}
\label{app:flow-and-filtering}

This appendix describes the flow representation given to the synthesis agent and the benign fingerprint library used to remove recurring background traffic before LLM processing.

\subsection{Protocol-Aware Flow Representation}
\label{app:protocol-fields}

The synthesis agent receives protocol-aware flow records rather than raw packets. Table~\ref{tab:protocol-fields} lists the per-protocol fields included in the
flow JSON after removing fields without Suricata equivalents and remapping
\texttt{tshark} names to Suricata-compatible keywords.

\begin{table*}[t]
\centering
\scriptsize
\caption{Per-protocol fields included in network flows.}
\label{tab:protocol-fields}
\setlength{\tabcolsep}{4pt}
\renewcommand{\arraystretch}{1.08}
\begin{tabular}{@{}c p{0.12\textwidth}p{0.78\textwidth}@{}}
\toprule
\textbf{\#} & \textbf{Protocol} & \textbf{Fields in flow JSON} \\
\midrule
1 & HTTP
& \texttt{http.request.method|http.method}, \texttt{http.host}, \texttt{http.request.full\_uri},
\texttt{http.request.line}, \texttt{http.request.version}, \texttt{http.user\_agent},
\texttt{http.accept}, \texttt{http.connection}, \texttt{http.content\_type},
\texttt{http.content\_length}, \texttt{http.response.code|http.stat\_code},
\texttt{http.response.line}, \texttt{http.server}, \texttt{http.file\_data},
\texttt{http.request.headers} parsed from \texttt{request\_line}. \\

2 & DNS
& \texttt{dns.qry.name}, \texttt{dns.qry.type}, \texttt{dns.qry.class},
\texttt{dns.flags.response}, \texttt{dns.flags.rcode|dns.rcode},
\texttt{dns.count.queries}, \texttt{dns.count.answers}, \texttt{dns.count.auth\_rr},
\texttt{dns.count.add\_rr}, \texttt{dns.resp.name}, \texttt{dns.a}, \texttt{dns.cname}. \\

3 & TLS
& \texttt{tls.handshake.extensions\_server\_name|tls.sni},
\texttt{tls.handshake.type}, \texttt{tls.record.length}, \texttt{tls.version}. \\

4 & TCP
& 5-tuple fields, \texttt{tcp.flags}, \texttt{tcp.flags.str},
\texttt{tcp.window\_size\_value}, \texttt{tcp.options.mss\_val},
\texttt{tcp.seq}, \texttt{tcp.ack}, \texttt{tcp.len}, \texttt{tcp.checksum},
\texttt{tcp.checksum.status}, and \texttt{tcp.payload.strings} for printable ASCII runs
$\geq 8$ characters, domains, and URI-like patterns. \\

5 & SMB
& \texttt{nbss.type}, \texttt{smb.cmd}, \texttt{smb.tid}, \texttt{smb.uid},
\texttt{smb.mid}, \texttt{smb.nt\_status}, \texttt{smb.path},
\texttt{smb.trans\_name}. \\

6 & NBDGM
& \texttt{nbdgm.type}, \texttt{nbdgm.src.ip}, \texttt{nbdgm.src.port},
\texttt{nbdgm.dgram\_len}, \texttt{nbdgm.source\_name},
\texttt{nbdgm.destination\_name}, \texttt{smb.server\_component},
\texttt{smb.cmd}, \texttt{smb.error\_class}, \texttt{smb.trans\_name},
\texttt{mailslot.opcode}, \texttt{mailslot.class}, \texttt{mailslot.name},
\texttt{browser.command}, \texttt{browser.period},
\texttt{browser.windows\_version}, \texttt{browser.os\_major},
\texttt{browser.server\_type}. \\

7 & LLMNR
& \texttt{dns.flags.response}, \texttt{dns.count.queries},
\texttt{dns.count.answers}, \texttt{dns.count.auth\_rr},
\texttt{dns.count.add\_rr}, \texttt{dns.qry.name}, \texttt{dns.qry.type},
\texttt{dns.qry.class}, \texttt{dns.retransmit\_request},
\texttt{\_ws.expert}, \texttt{\_ws.expert.severity}. \\
\midrule
\multicolumn{3}{@{}p{0.96\textwidth}@{}}{
\textit{Note:} All flows include the 5-tuple, transport protocol, and packet counts
(\texttt{packets\_total}, \texttt{packets\_src\_to\_dst}, and
\texttt{packets\_dst\_to\_src}).
} \\
\bottomrule
\end{tabular}
\end{table*}

\subsection{Benign Fingerprint Categories}
\label{app:fingerprint-categories}

The benign fingerprint library captures recurring identifiers observed in benign
executions, such as DNS names, HTTP hosts, destination IPs, and local name
resolution artifacts. Table~\ref{tab:benign-fingerprint-categories} summarizes
the fingerprint categories and representative filtering patterns.

\subsection{Filtering Impact}
\label{app:benign-filter-impact}

We measure the effect of benign fingerprinting on the malware corpus before LLM processing. Table~\ref{tab:benign-filter-impact} reports how many background and
security-relevant flows are removed, along with the reduction in flows passed to the LLM.

\begin{table}[t]
\centering
\scriptsize
\caption{Benign fingerprint categories and common filtering patterns.}
\label{tab:benign-fingerprint-categories}
\setlength{\tabcolsep}{3pt}
\renewcommand{\arraystretch}{1.08}
\begin{tabular}{@{}p{0.26\columnwidth}p{0.11\columnwidth}p{0.54\columnwidth}@{}}
\toprule
\textbf{Category} & \textbf{Count} & \textbf{Filtering Examples} \\
\midrule
DNS names
& 159
& Reverse-DNS lookups such as \texttt{in-addr.arpa} and \texttt{ip6.arpa}; routine Windows and system lookups. \\

HTTP hosts
& 37
& Common update and infrastructure hosts, including Microsoft, Apple, DigiCert, update, and time-service domains. \\

Destination IPs
& 49
& Repeated DNS, NTP, and operating-system service endpoints. \\

LLMNR names
& 27
& Local multicast name-resolution queries. \\

NBNS names
& 57
& Windows NetBIOS and file-sharing broadcasts. \\

NBDGM mailslots
& 5
& NetBIOS datagram destinations and mailslot names. \\

SSDP fingerprints
& N/A
& UPnP and service-discovery multicast traffic. \\
\bottomrule
\end{tabular}
\end{table}

Table~\ref{tab:benign-filter-impact} reports the effect of benign traffic fingerprinting on the 1,296 malware PCAPs used in the main evaluation. The filter removes most background flows while preserving nearly all security-relevant flows.

\begin{table}[t]
  \caption{Benign traffic fingerprinting.}
  \label{tab:benign-filter-impact}
  \small
  \begin{tabular}{@{}lr@{}}
    \toprule
    \textbf{Metric} & \textbf{Value} \\
    \midrule
    \multicolumn{2}{@{}l}{\textit{Flow composition (1,296 PCAPs)}} \\[2pt]
    Total raw flows                    & 131,787 \\
    Security-relevant flows            & 6,937 (5.3\%) \\
    Background flows                   & 124,850 (94.7\%) \\
    \midrule
    \multicolumn{2}{@{}l}{\textit{After benign fingerprinting}} \\[2pt]
    Security-relevant flows removed    & 142 (2.0\% of security-relevant) \\
    Background flows removed           & 110,879 (88.8\% of background) \\
    Flows passed to LLM                & 20,766 (15.8\% of total) \\
    Median flows/PCAP (raw)            & 65 \\
    Median flows/PCAP (filtered)       & 12 \\
    \bottomrule
  \end{tabular}
\end{table}

\section{Prompt Templates}
\label{app:prompts}

This appendix provides the prompt templates used by the Rule Synthesis Agent. Each template is filled at runtime with flow context, direction hints, and the serialized flow records.

\subsection{Zero-Shot Prompt}
\label{app:prompt-zeroshot}

The zero-shot setting gives the model only the task instructions and flow
records, without any worked examples. Figure~\ref{fig:prompt-zeroshot} shows the
zero-shot user prompt template.

\begin{figure}[!htbp]
\centering

\begin{tcblisting}{
  enhanced,
  width=\linewidth,
  colback=promptBody,
  colframe=promptBorder,
  coltitle=white,
  colbacktitle=promptHeader,
  title=\textbf{User Prompt --- Zero-Shot},
  fonttitle=\ttfamily\scriptsize,
  boxrule=0.5pt,
  arc=1pt,
  left=4pt,
  right=4pt,
  top=4pt,
  bottom=4pt,
  before skip=2pt,
  after skip=2pt,
  listing only,
  listing engine=listings,
  listing options={
    basicstyle=\ttfamily\tiny,
    breaklines=true,
    breakatwhitespace=false,
    columns=fullflexible,
    keepspaces=true,
    showstringspaces=false,
    frame=none,
    backgroundcolor=\color{promptBody},
    aboveskip=0pt,
    belowskip=0pt
  }
}
Total flows: {num_flows}

{flow_review_context}

Analyze the following network flows. Classify each flow and
generate Suricata rules as described in the system instructions.
Use only evidence present in the provided flow records.

{direction_hints}

OUTPUT FORMAT:
Return ONLY valid JSON -- no markdown, no prose.

{"classification":[{"flow_name":"flow_N","tag":"imp"}],
 "rules":["alert dns ... (msg:\"Rule\"; sid:...; ...)"]}

IMPORTANT: Only list flows you label "imp" in the classification
array. Flows you omit are treated as "unimp" automatically.

Flows JSON:
{flows_json}
\end{tcblisting}
\caption{Zero-shot user prompt template tokens are filled at runtime.}
\label{fig:prompt-zeroshot}
\end{figure}

\subsection{Chain-of-Thought Prompt}
\label{app:prompt-cot}

The chain-of-thought setting adds structured reasoning steps before the final JSON output, asking the model to identify protocol evidence, suspicious
indicators, traffic direction, and rule anchors. Figure~\ref{fig:prompt-cot}
shows the chain-of-thought prompt template.

\begin{figure}[!htbp]
\centering

\begin{tcblisting}{
  width=0.98\linewidth,
  colback=promptBody,
  colframe=promptBorder,
  coltitle=white,
  colbacktitle=promptHeader,
  title=\textbf{User Prompt --- Chain-of-Thought},
  fonttitle=\ttfamily\footnotesize,
  boxrule=0.6pt,
  arc=1pt,
  left=4pt,
  right=4pt,
  top=4pt,
  bottom=4pt,
  listing only,
  listing engine=listings,
  listing options={style=promptbox}
}
Total flows: {num_flows}

{flow_review_context}

Analyze the following network flows using chain-of-thought
reasoning.

{direction_hints}

STEP 1 -- CLASSIFICATION REASONING:
For each flow, think step-by-step:
  a) What protocol/application layer is this?
  b) What specific indicators are present?
  c) Are indicators suspicious in isolation or only combined?
  d) Does traffic show bidirectional exchange or one-sided probe?
  e) Final classification: imp or unimp? Why?

STEP 2 -- RULE GENERATION:
For each imp flow, use the mandatory rule headers from the system
prompt, then add detection keywords for the behavioral pattern.
Focus on generalizable patterns, not specific IOCs.

OUTPUT FORMAT:
Return ONLY valid JSON -- no markdown, no prose.

{
  "classification": [
    {
      "flow_name": "flow_N",
      "tag": "imp",
      "reason": "one-sentence summary of key evidence"
    }
  ],
  "rules": [
    "alert dns ... (msg:\"Rule\"; sid:...; ...)"
  ],
  "reasoning": "If zero rules, explain why."
}

Flows JSON:
{flows_json}
\end{tcblisting}

\caption{Chain-of-thought user prompt template.}
\label{fig:prompt-cot}
\end{figure}

\subsection{Few-Shot Prompt}
\label{app:prompt-fewshot}

The few-shot setting prepends worked flow-to-rule examples before the task
instructions. Figure~\ref{fig:prompt-fewshot} shows a representative example
that pairs a remapped flow record with its corresponding ET~Open rule.

\begin{figure}[!htbp]
\centering

\begin{tcblisting}{
  width=0.98\linewidth,
  colback=promptBody,
  colframe=promptBorder,
  coltitle=white,
  colbacktitle=promptHeader,
  title=\textbf{User Prompt --- Few-Shot Example},
  fonttitle=\ttfamily\footnotesize,
  boxrule=0.6pt,
  arc=1pt,
  left=4pt,
  right=4pt,
  top=4pt,
  bottom=4pt,
  listing only,
  listing engine=listings,
  listing options={style=promptbox}
}
--- EXAMPLE (HTTP Tier 1) ---
Flow:
{
  "app_protocol": "http",
  "src_ip": "192.168.168.206",
  "dest_ip": "178.32.110.193",
  "dest_port": 80,
  "packets_src_to_dst": 5,
  "packets_dst_to_src": 3,
  "data": {
    "http.method": "POST",
    "http.host": "prets-immobiliers.org",
    "http.uri": "/dbconnect.php",
    "http.user_agent": "Mozilla/5.0 (Windows NT 6.3; ...",
    "http.request_body": "data=EB7ED7C4C2C538751D5E01..."
  }
}

Classification: imp

Rule:
alert http $HOME_NET any -> $EXTERNAL_NET any
  (msg:"ET MALWARE Alphacrypt/TeslaCrypt Ransomware CnC Beacon";
   flow:established,to_server;
   http.method; content:"POST";
   http.uri; content:".php"; endswith;
   http.user_agent; content:"Mozilla"; startswith;
   http.request_body; content:"data="; startswith; fast_pattern;
   pcre:"/^[A-F0-9]{100,}$/R";
   classtype:trojan-activity; sid:2022504; rev:1;)

--- [examples 2-5 follow the same format] ---
\end{tcblisting}

\caption{Representative few-shot example, 1 of 5. Flow is shown post-remapping; rule is the matching ET~Open signature.}
\label{fig:prompt-fewshot}
\end{figure}

\section{Baseline Harness Prompt}
\label{app:harness-prompt}
This appendix reproduces the prompt given to the three agentic baselines of \S\ref{sec:setup}. OpenClaw and Hermes Agent read it from \texttt{AGENTS.md}. Claude Code receives it via \texttt{--append-system-prompt}. All 603 sandbox copies hash to one MD5, so the baselines differ only in scaffold and backbone, not instructions. What is held fixed is the task, evidence constraint, rule budget, and scoring rule. 

\subsection{Task Description}
\label{app:harness-task}

Each sandbox also carries the following \texttt{TASK.md}, together with
\texttt{run\_suricata.sh}, which replays Suricata over that sandbox's capture
only, and empty \texttt{rules/} and \texttt{work/} directories.
\begin{tcblisting}{
  enhanced,
  width=0.98\linewidth,
  colback=promptBody, colframe=promptBorder, coltitle=white,
  colbacktitle=promptHeader, title=\textbf{TASK.md},
  fonttitle=\ttfamily\scriptsize,
  boxrule=0.5pt, arc=1pt, left=4pt, right=4pt, top=4pt, bottom=4pt,
  before skip=4pt, after skip=6pt,
  listing only, listing engine=listings,
  listing options={
    basicstyle=\ttfamily\tiny, breaklines=true, breakatwhitespace=false,
    columns=fullflexible, keepspaces=true, showstringspaces=false,
    frame=none, backgroundcolor=\color{promptBody},
    aboveskip=0pt, belowskip=0pt
  }
}
# Task
`capture.pcap` in this directory is a network capture from a single host.
Write Suricata rules that detect the malicious/noteworthy activity in it, into
`rules/generated.rules`.
Everything you need is in this directory. See the system prompt for the method,
the constraints, and what counts as a good rule.
## What is here
| path | what it is |
|---|---|
| `capture.pcap` | the capture -- your only evidence |
| `run_suricata.sh` | `./run_suricata.sh <rules_file> [outdir]` -- test your rules against the capture |
| `rules/` | write `generated.rules` here |
| `work/` | scratch space; anything you like |
`tshark` is available for inspecting the capture.
\end{tcblisting}

\section{Comprehensive Error Patterns}
\label{app:comprehensive-error-patterns}

This appendix lists the verifier failure patterns used by the Repair Agent.
Failures are grouped into syntax errors, no-trigger failures, and benign
false positives.

\subsection{Syntax-Error Patterns}
\label{app:syntax-error-patterns}

Syntax-error patterns correspond to rules that Suricata cannot parse or load.
Table~\ref{tab:error-patterns-syntax} lists these errors, their root causes, and
the associated repair strategies.

\begin{table*}[t]
\centering
\scriptsize
\caption{Syntax-error patterns.}
\label{tab:error-patterns-syntax}
\setlength{\tabcolsep}{4pt}
\renewcommand{\arraystretch}{1.08}
\begin{tabular}{@{}c p{0.19\textwidth}p{0.34\textwidth}p{0.36\textwidth}@{}}
\toprule
\textbf{\#} & \textbf{Error Type} & \textbf{Root Cause} & \textbf{Fix Strategy} \\
\midrule
1 & UNKNOWN\_KEYWORD
& Keyword does not exist, is misspelled, or a tshark field is used directly
& Replace with the correct Suricata keyword, e.g., \texttt{http.request.method} $\rightarrow$ \texttt{http.method}. \\

2 & MALFORMED\_HEADER
& Invalid protocol or direction token in the rule header
& Use a valid Suricata protocol and the direction token \texttt{->}. \\

3 & PORT\_SYNTAX
& Invalid port format or reversed port tokens
& Use valid formats such as numeric ports, ranges, port lists, or \texttt{any}. \\

4 & MISSING\_REQUIRED\_FIELD
& Missing semicolon causes the parser to consume the next keyword
& Add \texttt{;} after every rule option. \\

5 & MISSING\_SEMICOLON
& Missing or misplaced semicolon in rule options
& Ensure every keyword option ends with \texttt{;}. \\

6 & CONTENT\_QUOTE\_ERROR
& Unterminated quote or malformed content string
& Use double quotes, escape internal quotes, or hex-escape special characters. \\

7 & PCRE\_ERROR
& Invalid PCRE syntax or unsupported modifiers
& Fix regex syntax, balance parentheses/classes, and avoid unsupported modifiers. \\

8 & EMPTY\_STICKY\_BUFFER
& Sticky buffer appears without a following \texttt{content:} or \texttt{pcre:}
& Ensure every sticky buffer is followed by at least one match condition. \\

9 & ENDSWITH\_MISUSE
& \texttt{endswith;} is used without preceding \texttt{content:}
& Move the content match into the sticky buffer before applying \texttt{endswith;}. \\

10 & APP\_LAYER\_MISMATCH
& Rule header protocol conflicts with the sticky-buffer family
& Align protocol header and buffer family, e.g., \texttt{alert dns} with DNS buffers. \\

11 & INVALID\_SID
& \texttt{sid} is missing, non-numeric, or zero-valued
& Set \texttt{sid} to a positive integer. \\

12 & INVALID\_REV
& \texttt{rev} is missing or non-numeric
& Set \texttt{rev} to a positive integer. \\

13 & INVALID\_FLOW
& \texttt{flow:} contains unsupported tokens
& Use valid tokens such as \texttt{established}, \texttt{to\_server}, or \texttt{to\_client}. \\

14 & DUPLICATE\_SID
& Rule contains multiple \texttt{sid} fields
& Preserve one valid \texttt{sid} and remove duplicates. \\

15 & DUPLICATE\_REV
& Rule contains multiple \texttt{rev} fields
& Preserve one valid \texttt{rev} and remove duplicates. \\

16 & THRESHOLD\_SYNTAX
& Malformed \texttt{threshold:} keyword
& Remove the entire \texttt{threshold:...;} clause. \\

17 & DETECTION\_FILTER\_SYNTAX
& Malformed \texttt{detection\_filter:} keyword
& Remove the entire \texttt{detection\_filter:...;} clause. \\

18 & BSIZE\_TOO\_LARGE
& Content length exceeds declared \texttt{bsize} value
& Remove \texttt{bsize:N;} or adjust it to a valid length. \\

19 & TLS\_VERSION\_MISUSE
& \texttt{tls.version} uses invalid decimal or quoted format
& Use canonical constants such as \texttt{tls12} or \texttt{tls13}. \\

20 & FAST\_PATTERN\_NO\_CONTENT
& \texttt{fast\_pattern;} appears without preceding \texttt{content:}
& Remove the stray \texttt{fast\_pattern;} keyword or place it after a content match. \\

21 & LEGACY\_HTTP\_KEYWORD
& Old Suricata 2.x HTTP post-match modifiers are used
& Convert to modern sticky-buffer syntax, e.g., \texttt{http\_uri} $\rightarrow$ \texttt{http.uri}. \\

22 & NOCASE\_NO\_CONTENT
& \texttt{nocase;} appears without preceding \texttt{content:} or \texttt{pcre:}
& Place \texttt{nocase;} immediately after the content or PCRE it modifies. \\

23 & DNS\_RCODE\_MISUSE
& \texttt{dns.rcode} uses invalid string or malformed value
& Use numeric values or uppercase constants such as \texttt{NXDOMAIN}. \\

24 & SSL\_VERSION\_INVALID
& Legacy SSL/TLS version field uses malformed value
& Prefer \texttt{tls.version} with constants such as \texttt{tls12}. \\

25 & DIRECTION\_CONFLICT
& Rule mixes request-direction and response-direction buffers
& Separate into two rules or align buffers with \texttt{to\_server}/\texttt{to\_client}. \\
\bottomrule
\end{tabular}
\end{table*}

\subsection{No-Trigger Patterns}
\label{app:no-trigger-patterns}

No-trigger patterns correspond to rules that pass syntax validation but fail to
alert on the source malware PCAP. Table~\ref{tab:error-patterns-no-trigger}
summarizes the main causes and repair strategies for these failures.

\begin{table*}[t]
\centering
\scriptsize
\caption{No-trigger error patterns.}
\label{tab:error-patterns-no-trigger}
\setlength{\tabcolsep}{4pt}
\renewcommand{\arraystretch}{1.08}
\begin{tabular}{@{}c p{0.21\textwidth}p{0.33\textwidth}p{0.35\textwidth}@{}}
\toprule
\textbf{\#} & \textbf{Error Type} & \textbf{Root Cause} & \textbf{Fix Strategy} \\
\midrule
1 & DIRECTION\_MISMATCH
& Rule direction does not match actual traffic direction
& Align C2 traffic as \texttt{\$HOME\_NET} $\rightarrow$ \texttt{\$EXTERNAL\_NET} with \texttt{to\_server}. \\

2 & DIRECTION\_MISMATCH\_EXT\_SRC
& \texttt{\$EXTERNAL\_NET} is source while flow uses \texttt{to\_server}
& Swap network variables so the client side is the source. \\

3 & DIRECTION\_MISMATCH\_INBOUND
& Outbound protocol is written as traffic going to \texttt{\$HOME\_NET}
& Reverse the direction so \texttt{\$HOME\_NET} is the source. \\

4 & PROTOCOL\_BUFFER\_MISMATCH
& Alert protocol does not match the observed application layer
& Use the correct alert protocol, e.g., HTTP $\rightarrow$ \texttt{alert http}. \\

5 & DUPLICATE\_STICKY\_BUFFER
& Same sticky buffer appears multiple times
& Merge all content or PCRE matches under one sticky-buffer instance. \\

6 & PCRE\_R\_ON\_DNS\_QUERY
& PCRE uses the relative \texttt{/R} flag with \texttt{dns.query}
& Remove the \texttt{/R} modifier from DNS query PCREs. \\

7 & GET\_WITH\_REQUEST\_BODY
& Rule matches GET method with \texttt{http.request\_body}
& Use \texttt{http.uri} for GET traffic or change the method to POST if appropriate. \\

8 & SMB\_ASCII\_ENCODING
& ASCII strings are used for SMB2 content matching
& Convert SMB strings to UTF-16LE hex encoding. \\

9 & SMB\_SHARE\_BUFFER
& \texttt{smb.share} buffer is empty for some tree-connect operations
& Use raw UTF-16LE hex content matching instead of \texttt{smb.share}. \\

10 & ENDSWITH\_ON\_HTTP\_URI
& \texttt{endswith;} on \texttt{http.uri} fails when URI has query parameters
& Remove \texttt{endswith;} and match the stable URI substring. \\

11 & OVERLY\_ANCHORED\_DNS\_PCRE
& DNS PCRE uses both start and end anchors too strictly
& Remove the leading anchor or match a stable domain suffix. \\

12 & DETECTION\_FILTER\_PRESENT
& \texttt{detection\_filter} or \texttt{threshold} requires multiple matches
& Remove the clause and write a single-packet detection rule. \\

13 & OVER\_SPECIFIC\_CONTENT
& Content string is too long or sample-specific
& Replace with the shortest stable substring that appears across target flows. \\

14 & REQUEST\_BODY\_NO\_METHOD
& \texttt{http.request\_body} is used without checking HTTP method
& Add \texttt{http.method; content:"POST";} or remove body matching for GET traffic. \\

15 & HTTP\_\allowbreak RESPONSE\_\allowbreak BUFFER\_\allowbreak TO\_\allowbreak SERVER
& Response buffers are used with \texttt{flow:to\_server}
& Flip to \texttt{to\_client} or replace response buffers with request-side fields. \\

16 & TLS\_CERT\_SUBJECT\_DIRECTION
& Certificate fields are matched with client-to-server direction
& Flip to \texttt{to\_client} when matching certificate subject or issuer fields. \\
\bottomrule
\end{tabular}
\end{table*}
\subsection{False-Positive Patterns}
\label{app:false-positive-patterns}

False-positive patterns correspond to rules that also trigger on benign traffic.
Table~\ref{tab:error-patterns-fp} shows the narrowing strategy used for this
failure class.

\begin{table*}[t]
\centering
\scriptsize
\caption{False-positive error patterns}
\label{tab:error-patterns-fp}
\setlength{\tabcolsep}{4pt}
\renewcommand{\arraystretch}{1.08}
\begin{tabular}{@{}c p{0.20\textwidth}p{0.34\textwidth}p{0.36\textwidth}@{}}
\toprule
\textbf{\#} & \textbf{Error Type} & \textbf{Root Cause} & \textbf{Fix Strategy} \\
\midrule
1 & FALSE\_POSITIVE
& Detection pattern is too broad and also matches benign traffic
& Tighten content with malware-specific anchors, add negated benign patterns, or narrow protocol/flow scope. Context-guided repair uses both benign and malware flow exemplars to narrow the pattern. \\
\bottomrule
\end{tabular}
\end{table*}

\section{Repair Knowledge Base}
\label{app:error-pattern-kb}

This appendix summarizes how the Repair Agent uses the error taxonomy from
Appendix~\ref{app:comprehensive-error-patterns}. Some failures are corrected
with deterministic rewrites before any LLM call, while the remaining failures
use LLM-guided repair with verifier feedback and flow context.

\subsection{Repair Strategy Coverage}
\label{app:repair-strategy-coverage}

Table~\ref{tab:error-patterns-summary} summarizes how the 42 error patterns are
split across deterministic rewrites and LLM-guided repair.

\begin{table*}[t]
\centering
\footnotesize
\caption{Error pattern distribution and repair strategy coverage.}
\label{tab:error-patterns-summary}
\setlength{\tabcolsep}{5pt}
\begin{tabular}{@{}lcccp{0.36\textwidth}@{}}
\toprule
\textbf{Category} & \textbf{Total} & \textbf{Deterministic} & \textbf{LLM-Guided} & \textbf{Primary Mode} \\
\midrule
SYNTAX\_ERROR
& 25 & 7 & 18
& Keyword substitution, option-format normalization, delimiter repair, and metadata-preserving syntax fixes. \\

NO\_TRIGGER
& 16 & 10 & 6
& Direction and buffer alignment, protocol correction, anchor relaxation, and context-grounded content revision. \\

FALSE\_POSITIVE
& 1 & 0 & 1
& Context-guided narrowing using benign-triggering flows and malware flows that should remain matched. \\
\midrule
\textbf{TOTAL}
& \textbf{42} & \textbf{17} & \textbf{25}
& Hybrid deterministic and LLM-guided repair. \\
\bottomrule
\end{tabular}
\end{table*}

\subsection{Deterministic Rewrites}
\label{app:deterministic-rewrites}

Deterministic rewrites handle structural failures that can be corrected without
LLM reasoning, such as keyword normalization, direction fixes, and sticky-buffer cleanup. Table~\ref{tab:deterministic-rewrites} lists the rewrite rules applied
before the LLM repair loop.

\begin{table*}[t]
\centering
\scriptsize
\caption{Deterministic rewrite rules applied by the Repair Agent.}
\label{tab:deterministic-rewrites}
\setlength{\tabcolsep}{4pt}
\renewcommand{\arraystretch}{1.08}
\begin{tabular}{@{}c p{0.21\textwidth}p{0.33\textwidth}p{0.35\textwidth}@{}}
\toprule
\textbf{\#} & \textbf{Error Type} & \textbf{Root Cause} & \textbf{Rewrite Action} \\
\midrule
1 & ENDSWITH\_MISUSE
& \texttt{endswith} or \texttt{startswith} is used incorrectly
& Rewrite as valid Suricata syntax, e.g., \texttt{content:"x"; endswith;}. \\

2 & DIRECTION\_CONFLICT
& Rule header direction conflicts with \texttt{flow:to\_server} or \texttt{flow:to\_client}
& Fix the rule direction so the header and flow option agree. \\

3 & DETECTION\_FILTER\_SYNTAX
& \texttt{detection\_filter} clause is malformed
& Strip the malformed \texttt{detection\_filter} clause. \\

4 & BSIZE\_TOO\_LARGE
& Declared \texttt{bsize} value is inconsistent with matched content length
& Adjust or remove the invalid \texttt{bsize} option. \\

5 & FAST\_PATTERN\_NO\_CONTENT
& \texttt{fast\_pattern} appears without a preceding \texttt{content} match
& Remove the orphaned \texttt{fast\_pattern} keyword. \\

6 & LEGACY\_HTTP\_KEYWORD
& Legacy Suricata HTTP modifiers are used instead of sticky buffers
& Replace legacy modifiers with modern sticky-buffer syntax, e.g., \texttt{http\_uri} $\rightarrow$ \texttt{http.uri;}. \\

7 & UNKNOWN\_KEYWORD
& Wireshark or \texttt{tshark} field name is used as a Suricata keyword
& Substitute with the corresponding valid Suricata keyword. \\

8 & DIRECTION\_MISMATCH\_EXT\_SRC
& \texttt{\$EXTERNAL\_NET} is source while flow uses \texttt{to\_server}
& Swap network variables so the client side is the source. \\

9 & DIRECTION\_MISMATCH\_INBOUND
& Outbound protocol is written as traffic going to \texttt{\$HOME\_NET}
& Reverse the direction so \texttt{\$HOME\_NET} is the source. \\

10 & DUPLICATE\_STICKY\_BUFFER
& Same sticky buffer appears multiple times
& Merge all content or PCRE matches under one sticky-buffer instance. \\

11 & PCRE\_R\_ON\_DNS\_QUERY
& PCRE uses the relative \texttt{/R} flag with \texttt{dns.query}
& Remove the \texttt{/R} modifier from DNS query PCREs. \\

12 & GET\_WITH\_REQUEST\_BODY
& Rule matches GET method with \texttt{http.request\_body}
& Remove \texttt{http.request\_body} matching for GET-only traffic. \\

13 & ENDSWITH\_ON\_HTTP\_URI
& \texttt{endswith;} on \texttt{http.uri} fails when URI has query parameters
& Remove \texttt{endswith;} and match the stable URI substring. \\

14 & OVERLY\_ANCHORED\_DNS\_PCRE
& DNS or TLS PCRE uses strict anchors that prevent matching observed traffic
& Remove overly strict anchors or match a stable domain suffix. \\

15 & DETECTION\_FILTER\_PRESENT
& \texttt{detection\_filter} requires multiple matches and prevents single-flow detection
& Remove the clause and write a single-packet detection rule. \\

16 & HTTP\_\allowbreak RESPONSE\_\allowbreak BUFFER\_\allowbreak TO\_\allowbreak SERVER
& Response buffers are used with \texttt{flow:to\_server}
& Flip to \texttt{to\_client} or replace response buffers with request-side fields. \\

17 & TLS\_CERT\_SUBJECT\_DIRECTION
& Certificate fields are matched with client-to-server direction
& Flip to \texttt{to\_client} when matching certificate subject or issuer fields. \\

\bottomrule
\end{tabular}
\end{table*}

\subsection{LLM-Guided Repairs}
\label{app:llm-guided-repairs}

LLM-guided repairs handle failures that require contextual reasoning over the
failed rule, verifier output, and relevant malware or benign flows.
Table~\ref{tab:llm-rewrites} lists the error patterns handled through these
context-aware repair prompts.

\begin{table*}[t]
\centering
\scriptsize
\caption{LLM-guided repair error patterns.}
\label{tab:llm-rewrites}
\setlength{\tabcolsep}{4pt}
\renewcommand{\arraystretch}{1.08}
\begin{tabular}{@{}c p{0.21\textwidth}p{0.33\textwidth}p{0.35\textwidth}@{}}
\toprule
\textbf{\#} & \textbf{Error Type} & \textbf{Root Cause} & \textbf{Fix Strategy} \\
\midrule
1 & MALFORMED\_HEADER
& Rule header contains an invalid protocol, direction token, or address/port structure
& Rewrite the header using a valid Suricata action, protocol, address, port, and direction. \\

2 & PORT\_SYNTAX
& Port value or port range is invalid or reversed
& Replace with a valid port, range, or \texttt{any} when the observed flow does not justify a fixed port. \\

3 & MISSING\_SEMICOLON
& Missing \texttt{;} causes the parser to merge or lose later rule options
& Add missing semicolons between rule options. \\

4 & CONTENT\_QUOTE\_ERROR
& Content string has an unterminated or malformed quote
& Rewrite the \texttt{content} option with balanced quotes and escaped special characters. \\

5 & PCRE\_ERROR
& PCRE contains invalid syntax, unbalanced groups, or unsupported modifiers
& Simplify or rewrite the PCRE using valid Suricata-compatible syntax. \\

6 & EMPTY\_STICKY\_BUFFER
& Sticky buffer appears without a following \texttt{content} or \texttt{pcre} match
& Add a valid match under the buffer or remove the unused sticky buffer. \\

7 & APP\_LAYER\_MISMATCH
& HTTP, DNS, TLS, or SMB buffer is used in a rule with the wrong application protocol
& Change the alert protocol or replace the buffer with one matching the observed application layer. \\

8 & INVALID\_SID
& \texttt{sid} is missing, non-numeric, duplicated, or zero-valued
& Replace with a single valid numeric \texttt{sid}. \\

9 & INVALID\_REV
& \texttt{rev} is missing or non-numeric
& Replace with a valid numeric \texttt{rev}. \\

10 & INVALID\_FLOW
& \texttt{flow:} contains an unrecognized or incompatible token
& Rewrite \texttt{flow:} using valid options such as \texttt{to\_server}, \texttt{to\_client}, and \texttt{established}. \\

11 & DUPLICATE\_SID
& Two rules share the same \texttt{sid}, or one rule contains multiple \texttt{sid} fields
& Keep one \texttt{sid} per rule and assign unique identifiers across generated rules. \\

12 & DUPLICATE\_REV
& \texttt{rev} appears more than once in one rule
& Keep a single \texttt{rev} field. \\

13 & THRESHOLD\_SYNTAX
& \texttt{threshold} clause is malformed
& Rewrite the clause using valid threshold syntax or remove it if not required for detection. \\

14 & OTHER\_SYNTAX
& Suricata parser error does not match a specific known pattern
& Use the parser feedback to check option ordering, semicolons, parentheses, and unsupported keywords. \\

15 & NOCASE\_NO\_CONTENT
& \texttt{nocase} appears without a preceding \texttt{content} option
& Move \texttt{nocase} after a valid \texttt{content} match or remove it. \\

16 & DNS\_RCODE\_MISUSE
& \texttt{dns.rcode} uses an invalid format or out-of-range value
& Replace with a valid DNS response code value or remove the condition. \\

17 & TLS\_VERSION\_MISUSE
& \texttt{tls.version} uses an invalid decimal or unsupported version string
& Use canonical Suricata constants such as \texttt{tls1.0} or \texttt{tls1.3}. \\

18 & SSL\_VERSION\_INVALID
& Legacy \texttt{ssl\_version} uses an unrecognized version constant
& Replace with a valid TLS/SSL version keyword or use the modern Suricata TLS version field. \\

19 & DIRECTION\_MISMATCH
& Rule direction or \texttt{flow:} does not match the observed traffic direction
& Align the rule direction with the observed client/server flow. \\

20 & PROTOCOL\_BUFFER\_MISMATCH
& Alert protocol does not match the observed application layer
& Use the correct alert protocol, e.g., HTTP $\rightarrow$ \texttt{alert http}. \\

21 & SMB\_ASCII\_ENCODING
& ASCII content is used for SMB2 traffic
& Convert SMB strings to UTF-16LE hex encoding. \\

22 & SMB\_SHARE\_BUFFER
& \texttt{smb.share} buffer is empty for tree-connect operations
& Use raw UTF-16LE hex content matching instead of \texttt{smb.share}. \\

23 & OVER\_SPECIFIC\_CONTENT
& Content string is too long or contains session-specific tokens
& Replace with a shorter stable substring that appears in the target flows. \\

24 & REQUEST\_BODY\_NO\_METHOD
& \texttt{http.request\_body} is used without checking the HTTP method
& Add \texttt{http.method; content:"POST";} or remove body matching for GET traffic. \\

25 & FALSE\_POSITIVE
& Rule also triggers on benign traffic
& Tighten the match with malware-specific content, add negated benign patterns, or narrow protocol and flow scope. \\

\bottomrule
\end{tabular}
\end{table*}

\section{Fixer Agent Feedback Templates}
\label{app:prompt-feedback}

This appendix shows the prompts used when a generated rule fails verification.
The Repair Agent uses one shared system prompt and three failure-specific
feedback templates for syntax failures, no-trigger failures, and benign false
positives.

\subsection{Fixer Agent System Prompt}
\label{app:fixer-system-prompt}

The shared system prompt defines the Repair Agent's role and constrains the
output to a corrected Suricata rule string. Figure~\ref{fig:prompt-fixer-system} shows the system prompt.

\begin{figure}[!htbp]
\centering
\begin{tcblisting}{
  width=0.98\linewidth,
  colback=promptBody,
  colframe=promptBorder,
  coltitle=white,
  colbacktitle=promptHeader,
  title=\textbf{Fixer Agent System Prompt},
  fonttitle=\ttfamily\footnotesize,
  boxrule=0.6pt,
  arc=1pt,
  left=4pt,
  right=4pt,
  top=4pt,
  bottom=4pt,
  listing only,
  listing engine=listings,
  listing options={style=promptbox}
}
You are a Suricata rule expert. Fix the provided broken rule
based on the error details.

Preserve all existing metadata fields. If metadata is missing,
add required ET metadata fields: affected_product, attack_target,
created_at, deployment, signature_severity, tag, updated_at.

Return ONLY the corrected rule string -- no markdown, no explanation.
\end{tcblisting}
\caption{Fixer-agent system prompt template.}
\label{fig:prompt-fixer-system}
\end{figure}
\FloatBarrier

\subsection{Syntax-Invalid Feedback Template}
\label{app:syntax-feedback-template}

For syntax failures, the verifier provides the failed rule, the Suricata parser error, and the matching knowledge-base guidance. Figure~\ref{fig:prompt-syntax-invalid-feedback}
shows the syntax-invalid feedback template.

\begin{figure}[!htbp]
\centering
\begin{tcblisting}{
  width=0.98\linewidth,
  colback=promptBody,
  colframe=promptBorder,
  coltitle=white,
  colbacktitle=promptHeader,
  title=\textbf{SYNTAX\_INVALID Feedback Template},
  fonttitle=\ttfamily\footnotesize,
  boxrule=0.6pt,
  arc=1pt,
  left=4pt,
  right=4pt,
  top=4pt,
  bottom=4pt,
  listing only,
  listing engine=listings,
  listing options={style=promptbox}
}
Rule to fix:
{rule_text}

Error/Problem:
{suricata_parser_error}

Knowledge base:
{knowledge_base_section}

Return ONLY the corrected rule string.
\end{tcblisting}
\caption{Syntax-invalid feedback prompt template. Runtime placeholders are filled with the failed rule, parser error, and matching syntax-error knowledge-base entry.}
\label{fig:prompt-syntax-invalid-feedback}
\end{figure}
\FloatBarrier

\subsection{No-Trigger Feedback Template}
\label{app:no-trigger-feedback-template}

For no-trigger failures, the rule passes syntax validation but does not alert on
the source malware PCAP. Figure~\ref{fig:prompt-no-trigger-feedback} shows the template used to inject verifier feedback and malware flow context.

\begin{figure}[!htbp]
\centering
\begin{tcblisting}{
  width=0.98\linewidth,
  colback=promptBody,
  colframe=promptBorder,
  coltitle=white,
  colbacktitle=promptHeader,
  title=\textbf{NO\_TRIGGER Feedback Template},
  fonttitle=\ttfamily\footnotesize,
  boxrule=0.6pt,
  arc=1pt,
  left=4pt,
  right=4pt,
  top=4pt,
  bottom=4pt,
  listing only,
  listing engine=listings,
  listing options={style=promptbox}
}
The following Suricata rule passed syntax check but did NOT
trigger on the malware PCAP:
{rule_text}

{sub_type_hint}

Knowledge base:
{knowledge_base_section}

Rule-vs-traffic analysis:
{flow_context}

Rewrite this rule so it triggers on the target malware traffic.

Constraints:
- Compare current anchors against observed traffic.
- Remove anchors absent from target traffic.
- Prefer anchors repeated across target flows.
- Avoid exact full URLs or brittle regexes unless supported.

Return ONLY the corrected rule string.
\end{tcblisting}
\caption{No-trigger feedback prompt template. Runtime placeholders are filled with the failed rule, no-trigger subtype, knowledge-base guidance, and malware flow context.}
\label{fig:prompt-no-trigger-feedback}
\end{figure}
\FloatBarrier

\subsection{False-Positive Feedback Template}
\label{app:false-positive-feedback-template}

For benign false positives, the verifier provides both benign-triggering flows
and malware flows that should remain matched. Figure~\ref{fig:prompt-false-positive-feedback}
shows the template used to narrow the rule while preserving malware detection.

\begin{figure}[!htbp]
\centering
\begin{tcblisting}{
  width=0.98\linewidth,
  colback=promptBody,
  colframe=promptBorder,
  coltitle=white,
  colbacktitle=promptHeader,
  title=\textbf{FALSE\_POSITIVE Feedback Template},
  fonttitle=\ttfamily\footnotesize,
  boxrule=0.6pt,
  arc=1pt,
  left=4pt,
  right=4pt,
  top=4pt,
  bottom=4pt,
  listing only,
  listing engine=listings,
  listing options={style=promptbox}
}
The following Suricata rule triggers on benign traffic:
{rule_text}

It triggered on {n_benign} benign PCAP(s).

Knowledge base:
{knowledge_base_section}

Benign traffic that caused false positives:
{benign_context}

Malware traffic the rule correctly matched:
{malware_flow_context}

Task: make this rule MORE SPECIFIC so it still triggers on
malware traffic but stops triggering on benign traffic.

Strategies:
- Add content present in malware but absent from benign flows.
- Use negated content to exclude benign patterns.
- Narrow PCRE anchors to avoid common benign strings.
- Add bsize constraints if malware queries have distinctive length.

Return ONLY the corrected rule string.
\end{tcblisting}
\caption{False-positive feedback prompt template. Runtime placeholders are filled with the failed rule, benign-triggering flows, malware-matching flows, and false-positive repair guidance.}
\label{fig:prompt-false-positive-feedback}
\end{figure}
\FloatBarrier

\section{Rule Family Generalization}
\label{app:family-generalization}

A rule that triggers on its source PCAP does not by itself show how specific
or general it is across the broader dataset. To measure this, we perform a
source--target generalization analysis. For each source PCAP, we replay its
associated rule set against every other PCAP in the dataset, producing one
source--target pair per target. We apply the same procedure to both the
ground-truth ET~Open rules and the \system{}-generated rules, allowing us to
compare their cross-PCAP behavior.

We label each source--target pair using the malware-family labels. A pair is
\emph{same-family} if the source and target share the same family label, and
\emph{other-family} otherwise. We exclude PCAPs with a \texttt{SINGLETON} family label, since singleton sources have no family-labeled targets. This leaves 1,281 non-singleton PCAPs. In the few-shot-5 run, \system{} produced valid rules for 1,074 out of 1,281
non-singleton source PCAPs. For a fair comparison, we restrict both ET~Open and \system{}'s generated rules to
this matched 1,074-source subset and replay each source rule set against the
remaining 1,280 non-singleton target PCAPs, excluding the source PCAP itself.

For each pair, trigger rate measures whether the source rule set fires at
least once on the target PCAP. Because triggering alone does not imply correct
security coverage, we also compute FAS precision, recall, and F1 by comparing
the flows alerted by the source rule set on the target PCAP with that target's
ground-truth alerted flows. All metrics are reported as micro-averages,
pooling raw flow counts across all pairs within each relation type, so that
families with more source--target pairs contribute proportionally more to the
aggregate results.

\begin{figure}[t]
    \centering
    \includegraphics[width=0.98\columnwidth]{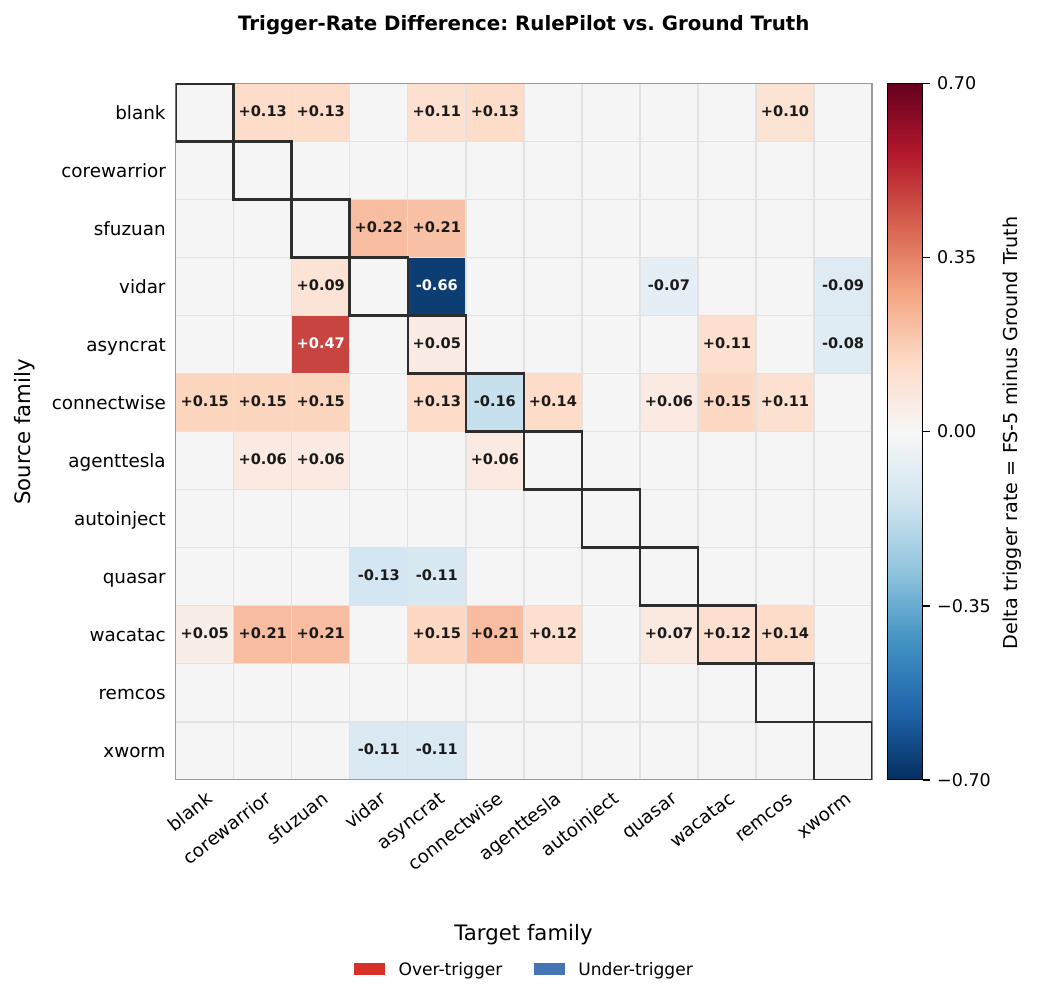}
    \caption{Trigger-rate difference between \system{} (few-shot-5) and ground-truth rules per source--target family pair.}
    \label{fig:generalizability}
\end{figure}

Table~\ref{tab:family-generalization} in \S\ref{sec:results-known} reports the aggregate same-family and other-family rates; overall, \system{}-generated rules exhibit \textit{generalization behavior} comparable to the ground-truth ET~Open rules, transferring to unseen same-family targets with near-ground-truth fidelity (FAS-F1: 0.67 vs.\ 0.75), consistent with the hypothesis that rules should align more strongly with samples from their own malware family than with samples from other families. Figure~\ref{fig:generalizability} below breaks this down further at the family-pair level.

Figure~\ref{fig:generalizability} visualizes trigger-rate differences across
the top-12 families. Red cells indicate target families where \system{}'s
generated rules over-trigger relative to ground truth; blue cells indicate
under-triggering. White cells indicate that \system{} matches
ground-truth triggering behavior on those family pairs. The largest deviations
are concentrated in a small number of source--target pairs: \emph{asyncrat}
$\rightarrow$ \emph{sfuzuan} shows strong over-triggering ($+0.47$), while
\emph{vidar} $\rightarrow$ \emph{asyncrat} shows strong under-triggering
($-0.66$). Several families, such as \emph{wacatac} and \emph{connectwise},
show mild over-triggering across multiple targets. For example,
(\emph{wacatac} $\rightarrow$ \emph{connectwise}) increases by $+0.21$, while
(\emph{connectwise} $\rightarrow$ \emph{wacatac}) increases by $+0.15$.
This suggests that cross-family leakage is concentrated in specific family
pairs rather than uniformly distributed across all pairs. Overall, \system{}
captures useful family-level patterns while largely maintaining cross-family
specificity.

\section{Additional Evaluation Results}
\label{app:additional-results}

This appendix reports supplementary evaluation results that support the main
evaluation but are omitted from the main text for space.

\subsection{Per-Variant Prompt Comparison}
\label{app:macro-prompt-comparison}

\begin{table*}[!t]
 \caption{Per-variant end-to-end performance across 1,296 malware PCAPs (full verification, \texttt{gpt-oss-120b}), macro-averaged.}
  \label{tab:appendix-full-end-to-end-performance}
  \small
  \centering
  \begin{tabular}{@{}lccc|cccccc@{}}
    \toprule
    & \multicolumn{3}{c|}{\textbf{Flow Classification}}
    & \multicolumn{6}{c}{\textbf{Rule Quality}} \\
    \cmidrule(lr){2-4}
    \cmidrule(lr){5-10}
    \textbf{Variant}
    & \textbf{P} & \textbf{R} & \textbf{F1}
    & \textbf{FAS-P} & \textbf{FAS-R} & \textbf{FAS-F1}
    & \textbf{Syntax} & \textbf{Trigger} & \textbf{FPR} \\
    \midrule
    Zero-shot   & 0.45 & 0.79 & 0.52 & 0.50 & 0.67 & 0.53 & 86.2\% & 80.9\% & 0.000\% \\
    CoT         & 0.44 & 0.77 & 0.51 & 0.50 & 0.64 & 0.52 & 84.5\% & 79.2\% & 0.006\% \\
    Few-shot-1  & 0.43 & 0.78 & 0.50 & 0.50 & 0.67 & 0.53 & 85.1\% & 80.5\% & 0.006\% \\
    Few-shot-3  & 0.43 & 0.78 & 0.51 & 0.51 & 0.68 & 0.53 & 86.1\% & 82.0\% & 0.000\% \\
    \textbf{Few-shot-5}
                & \textbf{0.44} & \textbf{0.79} & \textbf{0.52}
                & \textbf{0.51} & \textbf{0.68} & \textbf{0.54}
                & \textbf{85.3\%} & \textbf{81.1\%} & \textbf{0.006\%} \\
    CoT+FS-3    & 0.43 & 0.78 & 0.51 & 0.51 & 0.66 & 0.53 & 85.2\% & 80.0\% & 0.010\% \\
    \bottomrule
  \end{tabular}
\end{table*}

\subsection{Micro-Averaged Prompt Comparison}
\label{app:micro-prompt-comparison}

The main text reports macro-averaged metrics, where each PCAP contributes
equally. Table~\ref{tab:appendix-micro-end-to-end-performance} reports the
corresponding micro-averaged results, where true positives, false positives,
and false negatives are pooled across all 1,296 PCAPs before computing
precision, recall, and F1.

\begin{table*}[t]
  \caption{Micro-averaged end-to-end performance across 1,296 malware PCAPs.}
  \label{tab:appendix-micro-end-to-end-performance}
  \small
  \centering
  \begin{tabular}{@{}lccc|cccccc@{}}
    \toprule
    & \multicolumn{3}{c|}{\textbf{Flow Classification}}
    & \multicolumn{6}{c}{\textbf{Rule Quality}} \\
    \cmidrule(lr){2-4}
    \cmidrule(lr){5-10}
    \textbf{Variant}
    & \textbf{P} & \textbf{R} & \textbf{F1}
    & \textbf{FAS-P} & \textbf{FAS-R} & \textbf{FAS-F1}
    & \textbf{Syntax} & \textbf{Trigger} & \textbf{FPR (\%)} \\
    \midrule
    Zero-shot   & 0.422 & 0.870 & 0.568 & 0.502 & 0.749 & 0.601 & 94.40\% & 83.80\% & 0.0799 \\
    CoT         & 0.424 & 0.830 & 0.561 & 0.514 & 0.692 & 0.590 & 93.80\% & 82.80\% & 0.0721 \\
    Few-shot-1  & 0.410 & 0.869 & 0.557 & 0.503 & 0.766 & 0.608 & 93.90\% & 84.30\% & 0.0815 \\
    Few-shot-3  & 0.405 & 0.859 & 0.551 & 0.493 & 0.754 & 0.596 & 95.40\% & 84.80\% & 0.0819 \\
    Few-shot-5  & 0.417 & 0.868 & 0.563 & 0.505 & 0.771 & 0.611 & 95.10\% & 83.70\% & 0.0817 \\
    CoT+FS-3    & 0.411 & 0.840 & 0.552 & 0.514 & 0.735 & 0.605 & 93.70\% & 83.30\% & 0.0766 \\
    \bottomrule
  \end{tabular}
\end{table*}

\subsection{Repair Agent Outcome Details}
\label{app:repair-agent-outcomes}

Table~\ref{tab:repair-kb} breaks down Repair Agent outcomes by initial failure
status and knowledge-base section. It shows which failure types were most common
and which were most often repaired successfully.

\begin{table*}[t]
\centering
\caption{Repair Agent (Agent~3) knowledge-base sections accessed, GPT-OSS-120b full run.}
\label{tab:repair-kb}
\small
\begin{tabular}{llrrr}
\toprule
\textbf{Init.\ Status} & \textbf{KB Section} & \textbf{Rules} & \textbf{Fixed} & \textbf{Fix\%} \\
\midrule
  SYNTAX\_INVALID    & Unknown keyword                        &  40 &  15 &  38\% \\
                     & Other syntax error                     &  37 &   0 &   0\% \\
                     & Direction conflict                     &  35 &  27 &  77\% \\
                     & bsize value too large                  &   6 &   6 & 100\% \\
                     & TLS version keyword misuse             &   3 &   0 &   0\% \\
                     & Content quote error                    &   2 &   1 &  50\% \\
                     & DNS rcode misuse                       &   1 &   1 & 100\% \\
                     & Port syntax error                      &   1 &   0 &   0\% \\
                     & \hspace{1em}\textit{Subtotal} & 125 &  50 &  40\% \\
  \midrule
  NO\_TRIGGER        & No specific match (generic KB)         &  78 &  52 &  67\% \\
                     & endswith on http.uri                   &  10 &   4 &  40\% \\
                     & GET with request body                  &  10 &   8 &  80\% \\
                     & Direction mismatch                     &   9 &   8 &  89\% \\
                     & TLS cert.subject + direction           &   6 &   0 &   0\% \\
                     & HTTP resp. buffer on to-server         &   3 &   0 &   0\% \\
                     & Request body w/o method check          &   2 &   0 &   0\% \\
                     & SMB ASCII encoding                     &   1 &   1 & 100\% \\
                     & \hspace{1em}\textit{Subtotal} & 119 &  73 &  61\% \\
  \midrule
  FALSE\_POSITIVE    & Rule triggers on benign traffic        &   3 &   1 &  33\% \\
                     & \hspace{1em}\textit{Subtotal} &   3 &   1 &  33\% \\
  \midrule
\bottomrule
\end{tabular}
\end{table*}

\end{document}